\documentclass[reprint, amsmath,amssymb,prb]{revtex4-2}

\usepackage{graphicx}
\usepackage{amsmath}
\usepackage{natbib}
\usepackage{dcolumn}
\usepackage{bm}

\begin{document}

\title{Charge fractionalization beyond the Luttinger liquid paradigm: an analytical consideration}

\author{A. A. Dontsov}
\email{operatorne@yandex.ru}
\author{A. P. Dmitriev}
\affiliation{
Ioffe  Institute, Politekhnicheskaya 26,  194021,   St.~Petersburg,   Russia
}

\begin{abstract}
In this paper, we consider analytically the density evolution of a spinless Fermi liquid with a nonlinear dispersion relation into which one particle is injected. The interaction is point-like and the temperature is zero. We obtain a formula for the evolution of the density and discuss the picture it gives as well as the physics behind it. Compared to the case of a linear spectrum, we find further and more complex fractionalization of the initial density hump: it splits into three humps instead of two, moreover, all three change their shapes in a complicated manner. We analyze the mechanisms of these phenomena and calculate their main characteristics. We also show that the fractionalization can be illustrated from a semiclassical point of view.
\end{abstract}

\maketitle

\section{Introduction}

Over the last few years there has been considerable progress in the excitation and detection technology of single quasiparticles in 1-D systems that is of interest to basic research as well as practical applications. A photon in an optical waveguide is similar to a quasiparticle in a 1-D channel, so the latter systems might play the same role in terms of basic research as quantum optical systems do \cite{Bocquillon14,Grenier11}. The control of quasiparticles also allows one to work with quantum information \cite{Bauerle18}.

 \begin{figure}[]
\centerline{\includegraphics[width=1\linewidth]{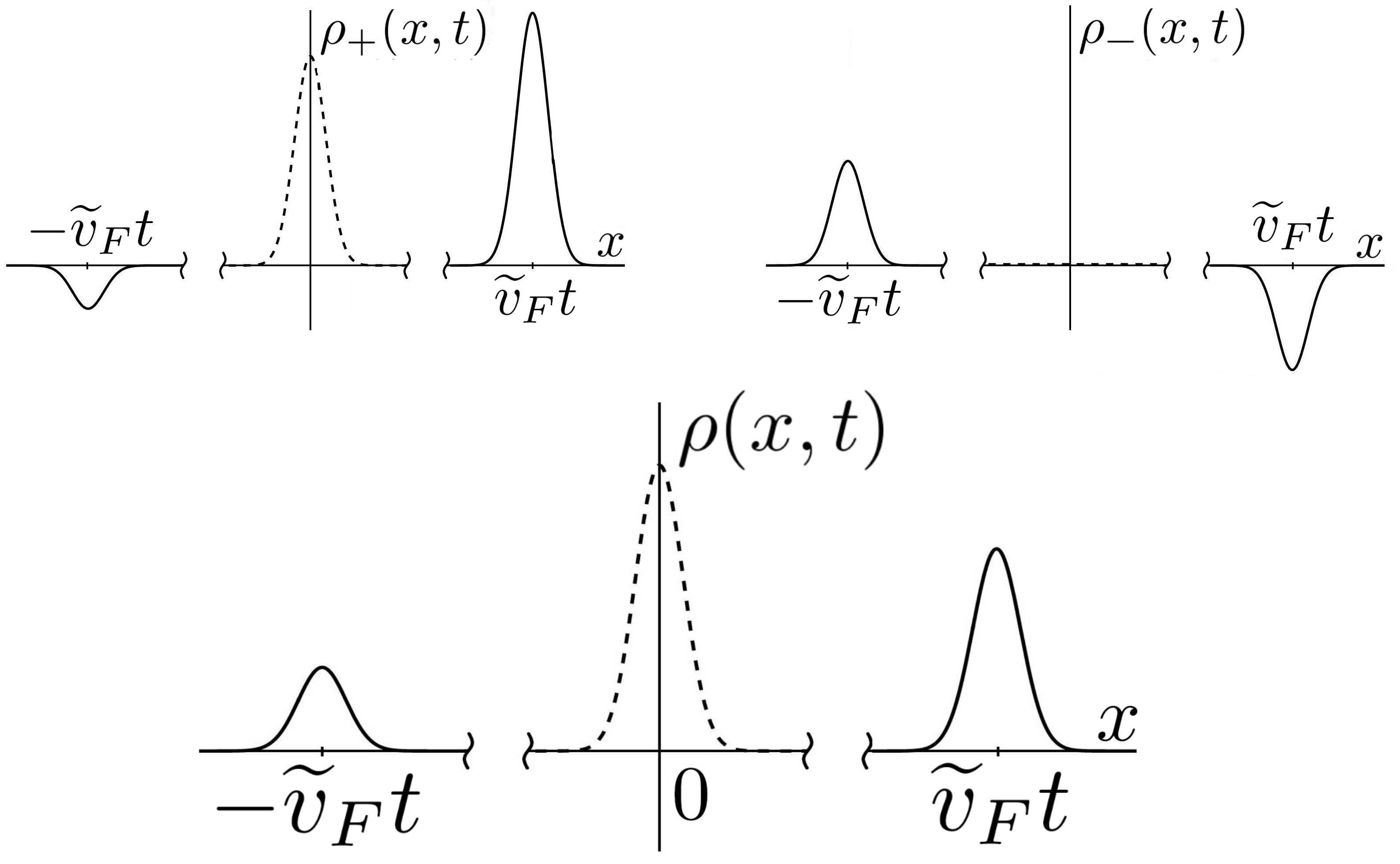}}
\caption{The density behavior of the right-moving $\rho_{+}(x,t)$ and the left-moving $\rho_{-}(x,t)$ particles when the dispersion relation is linear \cite{Das11}. The total density $\rho(x,t)=\rho_{+}(x,t)+\rho_{-}(x,t)$ is at the bottom. The dashed line shows the initial density distribution, the solid lines are the density when time $t$ is large enough for the opposite moving humps to split. The humps simply shift without changing their shapes.}
\label{fig:Lin}
\end{figure}

There is a diverse variety of 1-D systems: quantum wires, the chiral edge states of quantum Hall bars, the edge states of a two-dimensional topological insulator, carbon nanotubes etc. \cite{Deshpande10}. In most circumstances, these systems are well described by the Tomonaga-Luttinger (TL) model \cite{Deshpande10,Haldane81} that is applicable in the limit of low energies. This model is integrable and does not require using a perturbation theory. What is more, the naive perturbation approach leads to enormous mathematical difficulties. The reason is that  Luttinger quasiparticles are not fermions with renormalized properties, so the Landau-Fermi liquid theory cannot be used in one dimension \cite{Pham00}. 

Why the one-dimensional case is so exceptional can be illustrated by the following. Inject one electron into an interacting Luttinger liquid, the electron momentum distribution function is presumed to be concentrated near the right Fermi point; in real space $x$ it is concentrated around the origin forming a hump. The evolution of this density hump is quite unusual: it splits into two opposite moving parts as it is shown in fig. \ref{fig:Lin}.
 This effect is inherent in one-dimensional case, is called charge fractionalization \cite{Pham00,Steinberg08,Deshpande10, Das11} and has been explained theoretically in the limit of a linear dispersion relation by using the TL model \cite{Das11}. In addition, the model with spin 1/2 demonstrates so-called spin-charge fractionalization \cite{Hashisaka17,Giamarchi03}. In real experiments charge fractionalization is usually disguised by the interaction with edge electrode contacts, nevertheless, it can be detected by some indirect signs in non-DC experiments \cite{Calzona15}.  It has indeed been detected in quantum Hall samples \cite{Perfetto14_fracExpHall}, quantum wires \cite{Steinberg08}, and other systems.

The main TL model approximation is that the generic spectrum is replaced with a linear one. In the longest samples that have been created so far, the time during which a quasiparticle travels along the whole length is slight. So the nonlinearity of the dispersion relation cannot manifest itself within this time and, therefore, the TL model properly describes charge fractionalization in the situation where t$\rightarrow$0. Nevertheless, if it takes much longer for a quasiparticle to travel from one end to another, the nonlinearity becomes crucial. At large time scales, the simple behavior in fig. \ref{fig:Lin} gets much more complicated.

Excellent progress in the theory beyond the linear TL approximation has been achieved (see for review \cite{Imambekov12}), and the theory of composite fermions should be particularly highlighted \cite{Rozhkov05,Protopopov14}. In this paper, we apply these methods to consider analytically the problem of charge fractionalization unfolding over long timescales. Specifically, we will be calculating the density evolution of a system with a single injected electron when the propagation time can be significantly larger than it is usually considered. The nonlinearity of the spectrum that has to be taken into account results in remarkable density behavior. 

This problem has been numerically studied for the discrete t-J model \cite{Moreno13}, the further fractionalization was found, and the authors focused on spin-charge fractionalization. Here we investigate a system of spinless fermions with a quadratic dispersion relation, obtain analytical results and describe the fractions characteristics. We especially focus on the mechanisms that are behind the complicated spreading the fractions demonstrate. We too find further fractionalization of the initial density hump: in the case of a nonlinear spectrum, it splits into three humps instead of two. It is consistent with \cite{Moreno13}, except for, naturally, there is no spin fraction here. 

\begin{figure}[]
\centerline{\includegraphics[width=1\linewidth]{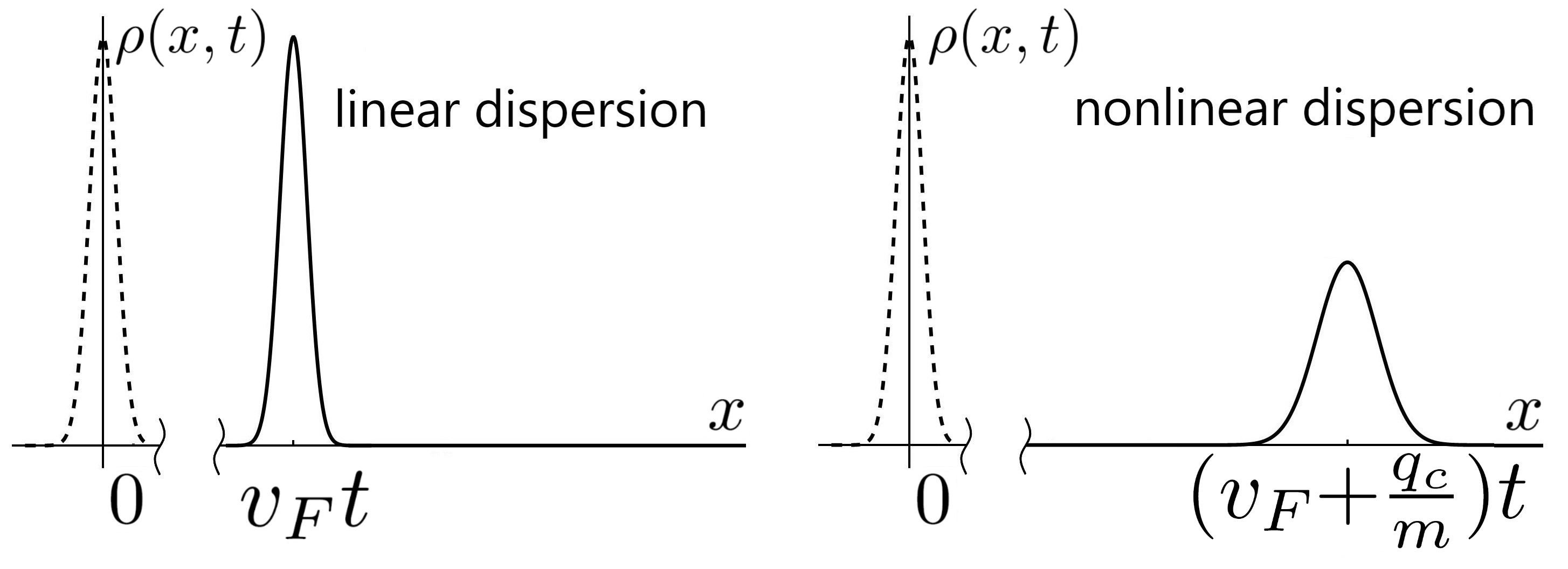}}
\caption{The density behavior of an injected particle in the absence of interaction. The particle momentum distribution is concentrated around $k_F+q_c$. The dashed line shows the initial density distribution, the solid lines show the distribution when the dispersion relation is linear (the group velocity is $v_F$) and nonlinear (the group velocity is $ v_F+\frac{q_c}{m}$ and the packet spreads out). The latter is the well-known solution of a single quantum particle problem.}
\label{fig:free}
\end{figure}

We also find that the shapes of the density humps change dramatically as time goes by. It is a stark contrast with the case of a linear spectrum, where the large-scaled density humps simply shift without changing their shapes (fig.\ref{fig:Lin}). The obvious phenomena that occurs is the simple spreading behavior a free particle wave packet demonstrates (fig.\ref{fig:free}), so, throughout the paper, we refer to this mechanism of spreading as ``free''. This familiar behavior is only pure in the absence of interaction. Otherwise, new unusual mechanisms of spreading discussed below take place, and the stronger the interaction becomes, the more significant these mechanisms are.

In order to make the picture bigger, we must mention another problem that is also under active consideration. The general problem is to examine what happens to classical hydrodynamic effects, notably to the shockwave effect \cite{Protopopov14,Veness19,Bettelheim12,Bettelheim06,Protopopov13}, in the quantum world. The problem of a shockwave involves analyzing of a density hump evolution, as ours does, and we would like to emphasize the difference, which is significant. 
The shockwave effect is closely associated with a “gradient catastrophe” regime:  if denser parts of a liquid move faster (or slower) than others, an initially smooth front develops large gradients \cite{Bettelheim06}. This requirement is fulfilled in many circumstances, however, it is irrelevant if the time of the gradient catastrophe $t_{grad}$ is much larger than the characteristic times of the effects one is interested in. Since the initial density deviation from equilibrium is slight in our case, $t_{grad}$ is, indeed, large; we will give an estimate of it at the end of the paper.

The paper is organized as follows. In the next Section \ref{sec:ProbStat} the problem is formulated in detail, and the main mathematical expression to be calculated is introduced. In Section \ref{sec:SemiclConsid}, the problem is considered from a semiclassical point of view, and it is shown that this simple approach, nevertheless, properly describes many features of the system. Even simpler semiclassical system that demonstrates the fractionalization effect is discussed in Appendix \ref{sec:SimEx}. Our quantum calculations are based on the theory of composite fermions, which is briefly discussed in Section \ref{sec:TheTheory}. In Section \ref{sec:LinQuantConsid} the methods of the theory are applied to the case of a linear dispersion relation, which solution was obtained earlier by using other methods. Section \ref{sec:NonLinQuantConsid} is the main one in the paper: there, we derive formulas for the density deviation from equilibrium in the case of a nonlinear spectrum; they are compared to known ones in some particular cases; the main characteristics of the density humps (amplitudes, velocities) are obtained from these formulas. We especially focus on the complicated mechanisms of changing the hump shapes.

\section{\label{sec:ProbStat}Problem statement}

 Let us consider a one-dimensional liquid at zero temperature that has a short-ranged interaction of a radius $a_0$. We inject a particle into the liquid and observe the density evolution. All deviations of the density are presumed to be large-scaled, i.e., theirs characteristic size $d$ is larger than both the Fermi wavelength $d>>\lambda_F$ and the interaction radius $d>>a_0$. That allows us to assume that the interaction is ``point-like'' \cite{Imambekov09} and $a_0\approx\lambda_F\approx \alpha$, where $\alpha$ is the parameter of the ``effective band-width'' \cite{Giamarchi03,vonDelft98}.

It should be emphasized again that the TL dispersion relation is linear, so the velocity of all the excitations is the same (in the limit of a large-scaled density deviation). It means that the density humps simply shift at the constant velocity (fig. \ref{fig:Lin}). This approximation is sufficient for small times, but as we are interested in what happens beyond that time frame, the nonlinearity has to be taken into account.

Denoting the ground state of the one-dimensional liquid by $|0\rangle$, we model the state of the liquid with an injected particle $|e\rangle$ by using the relation
\begin{eqnarray}
|e\rangle=\int\,dx\, \phi(x)\,\Psi^+(x) \,|0\rangle=\frac{1}{\sqrt{L}}\sum_{k=-\infty}^\infty \phi_k \,a^+_k \,|0\rangle ,
\end{eqnarray}
where $\phi_k=\int\,dx\, \phi(x) e^{-i x k}$, the operators $\Psi^+(x)$ and $a^+_k$ are the usual fermionic creation ones.

We presume that the ``wave function'' of the injected particle $\phi_k$ is concentrated around $k_F+q_c$ as it is shown in fig. \ref{fig:spec}. Its width is determined by the initial density hump characteristic size $d$ in real space $x$. The momentum distribution of the zero temperature ground state $P_{q}=\langle 0| a^+_{k_F+q} a_{k_F+q}|0\rangle$ is shown explicitly because it and its derivative will appear in the quantum calculations.

The goal of this paper is to examine in detail the density evolution within a relatively long time frame after an electron injection. The evolution can be expressed in terms of the density operator $\rho(x,t)$ as
\begin{eqnarray}
\label{eq:GenerRelat}
\langle e|\rho(x,t)|e \rangle=&&\langle\rho(x,t)\rangle_e\nonumber\\
=&&\iint \phi(x_2)\phi^*(x_1)\nonumber\\
&&\times \langle 0|\Psi(x_1) \rho(x,t)\Psi^+(x_2) |0 \rangle dx_1dx_2,
\end{eqnarray}
hereinafter the expected value ${\langle e| ...|e \rangle}$ is denoted by ${<...>_e}$, but ${\langle 0| ...|0 \rangle} ={ <...>}$. It can be seen that the problem mathematically comes down to calculating the average
\begin{equation}
\label{toCalculate}
\langle\Psi(x_1) \rho(x,t)\Psi^+(x_2)\rangle.
\end{equation}
\section{\label{sec:SemiclConsid} The semiclassical consideration in the case of a nonlinear dispersion relation}

Before proceeding to the quantum case, let us first consider the problem in a semiclassical formulation. Because of the exchange interaction, this semiclassical model cannot be considered to be the classical limit of the rigorous quantum model (more on that at the end of the section). However crude, it illustrates the main features the quantum consideration gives and helps tell what features are classical and what are quantum. Note that this semiclassical model is not the simplest one that demonstrates the fractionalization effect. Further simplifications can be made as we show in Appendix \ref{sec:SimEx}.

Our semiclassical consideration is based on classical Vlasov equation with omitted collision integral
\begin{equation}
\label{eq:Vlasov}
\frac{\partial f(x,p,t)}{\partial t}+\frac{p}{m}\frac{\partial f(x,p,t)}{\partial x}+F(x,t)\frac{\partial f(x,p,t)}{\partial p}=0,
\end{equation}
where $f(x,p,t)$ is a classical distribution function, $\int f(x,p,t)\frac{dp}{2 \pi}dx =N_T$, and $N_T$ is the total number of the particles in the system.  As usual, units are chosen so that $\hbar=1$. Then $F(x,t)=- \frac{\partial}{\partial x}\int  dx'  g(x-x') \rho^{(cl)}(x',t) $ is the averaged force acting on the liquid from liquid's density inhomogeneity, where  $g(x)$ is an interaction. And  $\rho^{(cl)}(x,t)$ is the density deviation, where the superscript 'cl' emphasizes that the approach is classical and $\rho^{(cl)}(x,t)$ is a function, not an operator.

 \begin{figure}[]
\centerline{\includegraphics[width=0.75\linewidth]{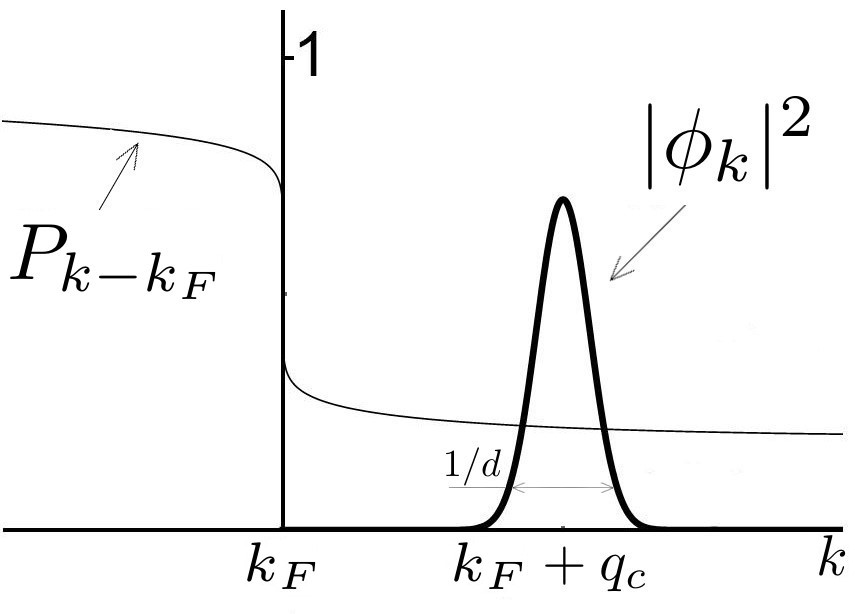}}
\caption{The momentum distribution of the zero temperature ground state $P_{q}=\langle 0| a^+_{k_F+q} a_{k_F+q}|0\rangle$ and of the injected particle $|\phi_k|^2$. We call the latter narrow when $\frac{1}{d}<<q_c$. Apart from that, we always presume $\frac{1}{d}<<k_F$.}
\label{fig:spec}
\end{figure}

We presume that $f(x,p,t)=f_0(p)+\delta f(x,p,t)$, where $\delta f(x,p,t)$ is a small deviation and $f_0(p)$ - the ``equilibrium'' distribution function. The density deviation then can be written as $\rho^{(cl)}(x,t)= \int  \frac{dp}{2 \pi} \delta f(x,p,t)$. For mathematical convenience's sake, choose the initial state in the form 
\begin{eqnarray}
	\delta f_q(p,0)=&&\int \delta f(x,p,0)e^{-i q x}dx \nonumber\\
	=&&\frac{2 \pi \hbar \sigma_p}{\pi[\sigma_p^2+(p-p_0)^2]}\,\rho^{(cl)}_q(0) ,
\end{eqnarray}
where $\rho^{(cl)}_q(0)$ is the initial density deviation that is assumed known; $p_0>m v_F$ is the average momentum of the particle injected into the system, and $\sigma_p$ is its spread in the momentum, which is presumed to be narrow for simplicity $\sigma_p<<p_0-p_F$. The characteristic size $d$ of the initial density distribution in real space $\rho^{(cl)}(x,0)$ is assumed to be much larger than the radius $a_0$ of the interaction $g(x)$, i.e., $d>>a_0$.

 Choosing the proper function $f_0(p)$ is essential to gain insight into what happens in quantum case. From mathematical point of view, any uniform function $f_0(p)$ is a solution of (\ref{eq:Vlasov}). That is because of omitting of the collision integral that is responsible for relaxation toward equilibrium. So one could think that it has to be the function shown in fig. \ref{fig:spec}. But the problem is that the equation (\ref{eq:Vlasov}) does not discriminate between equilibrium functions created by thermal processes at finite temperature like the Fermi-Dirac distribution and the functions created by interaction at zero temperature. The quantum consideration below will show, however, that the difference is significant. That is why, to avoid false predictions, we use the simple Fermi-Dirac function at zero temperature:
\begin{equation}
\label{eq:Fer-DirZerTemp}
f_0(p)=\theta(p_F-|p|).
\end{equation}

We do not describe the method of finding the solution here, it is demonstrated in detail in \cite{Pitaevskii2012}. The first order approximation is
\begin{eqnarray}
	\label{eq:VlasovSOL}
	\rho_q^{(cl)}(t)=&&   (\frac{ g_0 v_F }{2 \pi \widetilde v_F(\widetilde v_F-v_0)}) e^{-i q \widetilde v_F t}\,\, \rho^{(cl)}_q(0) \nonumber\\
	&&+ (\frac{ g_0 v_F}{2 \pi \widetilde v_F(\widetilde v_F+v_0)}) e^{+i q \widetilde v_F t}\,\, \rho^{(cl)}_q(0) \nonumber\\
	&& +(\frac{v_0^2-v_F^2}{v_0^2-\widetilde v_F^2}) e^{-i q v_0 t-\sigma_p \frac{|q|}{m} t}\,\rho^{(cl)}_q(0) ,
\end{eqnarray}
where $\widetilde v_F=v_F\sqrt{1+g_0/(\pi v_F)}$. 

Let us, for example, take the initial density deviation in the form $\rho^{(cl)}(x,0)=d/(x^2+d^2)$, in this case ${\rho^{(cl)}_q(0)=\pi e^{- |q| d}}$ and
\begin{eqnarray}
	\label{eq:VlasovSOLinX}
	\rho^{(cl)}(x,t)=&&\frac{1}{2 \pi}\int dq\, \rho_q^{(cl)}(t) e^{i q x} \nonumber \\
	=&&(\frac{ g_0 v_F }{2 \pi \widetilde v_F(\widetilde v_F-v_0)}) \frac{d}{(x-\widetilde v_F t)^2+d^2}\nonumber\\
	&&+(\frac{ g_0 v_F}{2 \pi \widetilde v_F(\widetilde v_F+v_0)})\frac{d}{(x+\widetilde v_F t)^2+d^2} \nonumber\\
	&&+(\frac{v_0^2-v_F^2}{v_0^2-\widetilde v_F^2})\frac{(d+\frac{\sigma_p}{m} t)}{(x- v_0 t)^2+(d+\frac{\sigma_p}{m} t)^2}
\end{eqnarray}

Note that the solution (\ref{eq:VlasovSOLinX}) properties discussed below do not depend on the initial hump shape $\rho^{(cl)}(x,0)$ and can be obtained directly from (\ref{eq:VlasovSOL}). In $x$ space, for instance, the first two terms in (\ref{eq:VlasovSOLinX}) are the initial humps moving at $\pm \widetilde v_F$ without changing their shapes. But it is a simple feature of the Fourier transform that if a function $\zeta_q(t)$ has the form  $\zeta_q(t)=\zeta_q(0) e^{-i q \widetilde v_F t}$, then 
\begin{equation}
	\label{eq:FurierTransfShift}
	\zeta(x,t)=\frac{1}{2\pi}\int dq \, \zeta_q(0) e^{i q (x-\widetilde v_F t)}=\zeta(x-\widetilde v_F t,0).
\end{equation}

If one takes the limit $m\rightarrow \infty$, then the spectrum becomes linear, and all the velocities become equal $v_0\rightarrow v_F$. The third term in (\ref{eq:VlasovSOLinX}) vanishes and, surprisingly enough, the solution (\ref{eq:VlasovSOLinX}) exactly coincides with the well-known \cite{Pham00,Steinberg08,Deshpande10, Das11} quantum solution of TL model, i.e., there are two humps moving at $\pm \widetilde v_F$ with the amplitudes $(1\pm K)/2$, where $K=1/\sqrt{1+g_0/(\pi v_F)}$ (see fig. \ref{fig:Lin}). Note that the ``quantum-like'' distribution function is the only quantum feature needed to describe charge fractionalization. So, one can see that for a linear dispersion relation, classical mechanics is enough to describe the effect.

  Although in the case of a linear spectrum, the semiclassical and the quantum considerations give the same result, when the spectrum is nonlinear, the results differ, but share some common features.
   In semiclassical case (\ref{eq:VlasovSOLinX}), when the spectrum is nonlinear, the first thing to notice is splitting of the right-moving hump into two humps moving at $\widetilde v_F$ and $v_0$. The latter corresponds to the injected particle (we call it ``fast'' hump) and the former to the plasmons of the liquid. The left-moving hump does not split and propagates at $-\widetilde v_F$. It can also be seen that the slow humps, again, do not change their shapes. They would, however, spread linearly in time, if the temperature was not zero, but we consider the situation of zero temperature only. The fast hump spreads as  $\Delta x_{fast}(t)\sim\frac{\sigma_p}{m} t$ when time $t$ is large. It is the usual free spreading typical in quantum mechanics (fig. \ref{fig:free}), except for ``the momentum uncertainty'' is due to the classical momentum spread $\sigma_p$. It should be stressed again that this behavior does not depend on the initial density distribution and can be inferred directly from the general solution (\ref{eq:VlasovSOL}). Table \ref{tab:CharsClassic} summarizes all the results.

  Note that in the semiclassical solution it is possible that the ``fast'' hump velocity is lesser than the one of the ``slow'' hump $v_0< \widetilde v_F$, in the quantum case it is not possible, so the name  ``fast''  is suitable.

    The goal of this section was to illustrate the fractionalization effect from a simple semiclassical point of view. As we will see, the picture is fairly similar in the quantum case: there will still be three humps and the free spreading mechanism still take place, but most of the hump characteristics such as velocity will change, the amplitudes may even change their sign and new mechanisms of quantum spreading will occur.
   
   That is why it should be emphasized that this crude model is not the classical limit of the rigorous quantum model considered below. We do not discuss the complicated \cite{home2013conceptual} classical limit problem here, but would like to stress the importance of the exchange interaction for our system, which does not allow one to take the classical limit at all \cite{davidson2017semiclassical}. Since the temperature is zero, the exchange interaction may play significant role in macroscopic systems leading, for example, to so called degeneracy pressure \cite{Shukla2011}. It is known that in some cases (in the mean-field regime), the exchange interaction can be neglected and then the quantum dynamic equation can be reduced to the Vlasov equation (\ref{eq:Vlasov}) by taking the classical thermodynamic limit $N_T\rightarrow\infty$ \cite{Benedikter2016,Benedikter2015}. In the general case, systems in coherent states can be reduced to classical ones by taking $\hbar\rightarrow 0$ \cite{Hepp1974}. However, the evolution of a general fermionic system in a general state cannot be described by any classical equations because of the exchange interaction \cite{davidson2017semiclassical}. This is the case for our process: the exchange interaction does play significant role and the quantum system equations cannot be reduced to (\ref{eq:Vlasov}) or any classical ones. Interestingly, the exchange interaction cannot manifest itself within a small time interval and the semiclassical model can describe the linear TL model as it was shown above.
   
    Nevertheless, our semiclassical consideration can be further refined by means of the Wigner equation without collision integral that takes the exchange interaction into account to some extent (see for example \cite{brodin2016quantum}). In this case, the velocities become closer to that of quantum case and the amplitudes have the right sign.



\begin{table}[]
\caption{\label{tab:CharsClassic}
The humps characteristics obtained by an analysis of the semiclassical solution (\ref{eq:VlasovSOL})
 }
\begin{ruledtabular}
\begin{tabular}{c r c }
	      Hump        & Velocity          & Spreading      \\ \hline
	      Fast        & $v_0$             & Free spreading \footnotemark[1] \\
	Right-moving slow & $\widetilde v_F$  & No spreading \footnotemark[2]   \\
	Left-moving slow  & $-\widetilde v_F$ & No spreading \footnotemark[2]
\end{tabular}
\end{ruledtabular}
\footnotetext[1]{The simple spreading behavior a free particle wave packet demonstrates (fig. \ref{fig:free}). In classical mechanics the momentum uncertainty is due to the classical momentum spread $\sigma_p$, while in quantum mechanics it is determined by Heisenberg uncertainty principle and equals to $1/d$}
\footnotetext[2]{The slow humps do not spread if the temperature is zero and the interaction is ``point-like'', otherwise they spread linearly in time.}
\end{table}

\section{\label{sec:TheTheory} The theory of composite fermions}

If the dispersion relation is nonlinear, it is rather difficult to calculate (\ref{toCalculate}), since $\rho(x,t)$ depends on time in a complicated sort of way. In order to find this dependence, use the theory of composite fermions \cite{Rozhkov05,Imambekov09,Protopopov14}.  

As usual, it is presumed that $\Psi(x)=\Psi_+(x)e^{i k_F x}+\Psi_-(x)e^{-i k_F x}$; $\Psi_\eta(x)=\frac{1}{\sqrt{L}}\sum_{q_1} c_{\eta,q_1} e^{i q_1 x}$; the subscripts $\eta=\pm 1$ denote right or left direction respectively and $c_{\eta,q_1}=a_{\eta \cdot k_F+q_1}$. The Hamiltonian of the nonlinear system up to the first order of $\frac{1}{m}$ reads \cite{Protopopov14}
\begin{eqnarray}
\label{HamiObi4}
 \widehat H = && \int \{ \pi v_F[\rho_+^2(x)+\rho_{-}^2(x)]+\frac{2 \pi^2}{3m}[\rho_+^3(x)+\rho_{-}^3(x)] \} dx\nonumber\\
&&+\frac{1}{2}\int g(x_1-x_2)\rho(x_1) \rho(x_2) dx_1 dx_2,
\end{eqnarray}
where $\rho(x)=\rho_+(x)+\rho_{-}(x)$.

Use Bogolubov transformation
\begin{equation}
\label{BogTran}
\rho_{\eta,q}=\cosh{(\theta_q)} \widetilde{\rho}_{\eta,q}-\sinh(\theta_q) \widetilde{\rho}_{-\eta,q},
\end{equation}
where $\tanh2\theta_q=\frac{g_q}{g_q+2 \pi v_F}$, then the Hamiltonian is recast into bosonic representation with cubic terms of $\widetilde{\rho}_{\eta,q}$. This Hamiltonian in turn can be rewritten in terms of new quasiparticles called ``composite fermions'' \cite{Rozhkov05}. The density of these new quasiparticles is $\widetilde{\rho}_{\eta,q}$ and using (\ref{BogTran}) the other particle operators can be defined. Although the relation (\ref{BogTran}) between the densities is simple, the one between field operators is much more complicated: 
\begin{equation}
	\Psi_\eta(x)=\widetilde{\Psi}_\eta(x)\,\widetilde{F}_\eta(x);
\end{equation}
\begin{equation}
	\label{Fvir}
	\widetilde{F}^+_\eta(x)=exp[-\frac{ 2 \pi \eta}{L}\sum_{q\neq0}\frac{e^{i q x}}{q}(w_q \widetilde{\rho}_{\eta,q}+u_q \widetilde{\rho}_{-\eta,q})],
\end{equation}
where $w_q=\cosh\theta_q-1$; $u_q=\sinh\theta_q$.

The refermionized Hamiltonian (\ref{HamiObi4}) has the form
\begin{equation}
\label{HamiKvasi}
\widehat H=\sum_{\eta, q_1} (\eta \widetilde v_F q_1+\frac{q_1^2}{2 m^*})\widetilde c_{\eta,q_1}^+ \widetilde c_{\eta,q_1}+`...`,
\end{equation}
where $\widetilde v_F=v_F\sqrt{1+{g_0}/{(\pi v_F)}}$; $m^*=m/(\cosh^3\theta_0-\sinh^3\theta_0)$; the symbol `...` means three terms that describe an interaction between the quasiparticles:  a term quadratic in $\widetilde{\rho}_{\eta,q}$, a cubic one, and a term that describes a ``point-like'' interaction between quasiparticles on opposite branches \cite{Rozhkov05,Imambekov09,Imambekov12,Protopopov14}. The main advantage of the Hamiltonian (\ref{HamiKvasi}) is that perturbation theory for the quasiparticles is regular \cite{Rozhkov05}, for our goals, however, it is enough to use the free part only, so that the terms denoted by the symbol `...` are omitted. The quadratic and cubic terms can be neglected if the interaction is point-like \cite{Imambekov09,Protopopov14}; the ``point-like'' interaction between the opposite branches can be neglected since the characteristic size of the density deviation is large $d>>\lambda_F$ \cite{Imambekov12}. In the end, the Hamiltonian reduces to a free one
\begin{equation}
\label{eq:free}
\widehat H=\sum_{\eta, q_1}  (\eta \widetilde v_F q_1+\frac{q_1^2}{2 m^*}) \widetilde c_{\eta,q_1}^+\widetilde c_{\eta,q_1}.
\end{equation}

Now when the Hamiltonian of the composite fermions is free, the time dependence of the composite fermion density operator is easy to find
 \begin{equation}
\label{easyrhoT}
 \widetilde{\rho}_{\eta,q}(t)=\sum_{q_1} \widetilde c_{\eta,q_1-q}^+\,\widetilde c_{\eta,q_1} \,e^{-i(\eta q\,\widetilde{v}_F-\frac{q^2}{2m^*}+\frac{q_{1}\,q}{m^*})t}.
 \end{equation}

There is a simple relation between the total densities for a point-like interaction ($\theta_q \approx \theta_0$), which follows from (\ref{BogTran})
 \begin{eqnarray}
\label{eq:rhoProport}
 \rho_{+,q}+\rho_{-,q}=&&(\cosh\theta_0-\sinh\theta_0)(\widetilde{\rho}_{+,q}+\widetilde{\rho}_{-,q}) \nonumber\\
 =&&\sqrt{K}(\widetilde{\rho}_{+,q}+\widetilde{\rho}_{-,q}),
 \end{eqnarray}
where the parameter $K$ is usually defined as 
\begin{equation*}
	K=1/\sqrt{1+\frac{g_0}{\pi v_F}}
\end{equation*}
and
\begin{equation*}
	\sinh{(\theta_0)}=\frac{1}{2}(\frac{1}{\sqrt{K}}-\sqrt{K}), \qquad \cosh{(\theta_0)}=\frac{1}{2}(\frac{1}{\sqrt{K}}+\sqrt{K}).
\end{equation*}

\section{\label{sec:LinQuantConsid} The quantum consideration in the case of a linear dispersion relation}

The theory described allows one to find easily the density evolution when the dispersion relation is linear, i.e., within TL approximations. This calculation is so straightforward that an analysis of (\ref{toCalculate}) is not needed. If $m^*\rightarrow\infty$, then (\ref{easyrhoT}) takes the form $\widetilde{\rho}_{\eta,q}(t)=\sum_{q_1} \widetilde c_{\eta,q_1-q}^+\,\widetilde c_{\eta,q_1} \,e^{-i \eta\, q\,\widetilde{v}_F t}=\widetilde{\rho}_{\eta,q} e^{-i \eta\, q\,\widetilde{v}_F t}$, so using  (\ref{BogTran}) we obtain
 \begin{subequations}
 \begin{eqnarray}
\label{LinstrogRight}
\rho_{+,q}(t)=&&\cosh^2(\theta_0) \rho_{+,q} e^{-i  q\,\widetilde{v}_F t} -\sinh^2(\theta_0) \rho_{+,q} e^{i  q\,\widetilde{v}_F t}\nonumber\\
&&+(...),
\end{eqnarray}
\begin{eqnarray}
\label{LinAns}
\rho_{-,q}(t)=&&\cosh(\theta_0)\sinh(\theta_0) (-\rho_{+,q} e^{-i  q\,\widetilde{v}_F t}+ \rho_{+,q} e^{i  q\,\widetilde{v}_F t})\nonumber\\
&&+(...),
\end{eqnarray}
\end{subequations}
 where the interaction is presumed to be point-like, so that $\theta_q=\theta_0$, the brackets (...) denote the terms depending on $\rho_{-,q}$ only. When averaged, these terms vanish. Make the transformation to the coordinate representation and assume that the initial densities are known $\langle\rho_{+}(x,0)\rangle_e=\rho_{+}^{(0)}(x)$ and $\langle\rho_{-}(x,0)\rangle_e=0$. The averaged density reads

 \begin{subequations}
 \label{RHO_Lin}
 \begin{eqnarray}
 \label{RHO_r_Lin}
\langle\rho_{+}(x,t)\rangle_e=&&\cosh^2(\theta_0)\rho_{+}^{(0)}(x-\widetilde{v}_F t)\nonumber\\
&&-\sinh^2(\theta_0)\rho_{+}^{(0)}(x+\widetilde{v}_F t),
\end{eqnarray}
\begin{eqnarray}
\langle\rho_{-}(x,t)\rangle_e=&&\cosh(\theta_0)\sinh(\theta_0)(-\rho_{+}^{(0)}(x-\widetilde{v}_F t)\nonumber\\
&&+\rho_{+}^{(0)}(x+\widetilde{v}_F t)),
\end{eqnarray}
\end{subequations}
where $\langle...\rangle_e$ means averaging over a state with an arbitrary initial density distribution $\rho_{+}^{(0)}(x)$.

We have again obtained the known \cite{Pham00,Steinberg08,Deshpande10, Das11} solution: two density humps with the amplitudes  $(1\pm K)/2$ that shift without changing their shapes (fig. \ref{fig:Lin}). 

What is most notable about this case is that the excitations of right- and left-moving particles are always comoving, and ones cannot exist without the others if there is any interaction. To form a relatively stable packet, the sum of them is required. That is the reason why the initial density hump involving only the right-moving particles then splits. And that is why the ratios of the amplitudes are the same 
$$\frac{\cosh^2(\theta_0)}{\cosh(\theta_0)\sinh(\theta_0)}=\frac{\cosh(\theta_0)\sinh(\theta_0)}{\sinh^2(\theta_0)}.$$

The second feature of the propagation is the absence of spreading. As it was mentioned before, for a linear spectrum and a large-scaled density deviation, all the excitation velocities are equal to $\pm \widetilde v_F$.

\section{\label{sec:NonLinQuantConsid}The quantum consideration in the case of a nonlinear dispersion relation}
For a nonlinear dispersion relation as well as for a linear one, the density humps of opposite chirality always comove. However, the shape evolution turns out to be significantly more complicated, as it can be seen from the rough semiclassical approximation (\ref{eq:VlasovSOL}). The quantum result is even more profound. We will see that most of the humps characteristic (velocity, amplitudes, etc) are different from that in the semiclassical case, but more importantly, both slow and fast humps experience an additional spreading that has quantum nature.

\subsection{\label{sec:ExcPrinc} The role of the exclusion principle}
By injecting a particle into the liquid, it is not possible to create an arbitrary initial density hump with an arbitrary local velocity distribution. It is because of the exclusion principle, and the relation (\ref{eq:GenerRelat}) reflects this fact as we will see from the quantum calculations below. But before doing the calculations, let us first discuss the physical meaning of it. As usual, we assume that the particle is injected relatively close to the right Fermi point. 

Note that the exclusion principle was not so important in the case of a linear spectrum, since the density behavior did not depend on momentum distribution as we saw in (\ref{RHO_Lin}); however, the reasoning below are applicable to the linear spectrum as well.

Consider the relation (\ref{eq:GenerRelat}). If $|0\rangle$ is a state of no particles, then the one particle in $|e \rangle$ has a wave function  $\phi(x)$ with the characteristic size $d$, in this case, the density is always $|\phi(x)|^2$. In momentum space, the wave function is written as $\phi_k=\int\phi(x) e^{-i k x}dx$, which width is $\frac{1}{d}$, and the center of which (fig. \ref{fig:spec}) is denoted as $k_F+q_c$, in the case of no particles, however, $k_F$ is just an arbitrary constant. 

Next, if $|0\rangle$ is the ground state of the free Fermi gas that has zero temperature, then an electron cannot be injected under the Fermi level $|k|<k_F$ because of the Exclusion Principle. So $\phi_k$ is generally cut, and, consequently, the density deviation may differ from $|\phi(x)|^2$.

Finally, if  $|0\rangle$ is a ground state of the interacting liquid, the state with the momentum $k_F+q$ is unoccupied with a probability of $1-P_q$, where $P_q=\langle 0| a^+_{k_F+q} a_{k_F+q}|0\rangle$ is the momentum distribution of the ground state. It means that an electron can only be partially (with a certain probability) injected into it. So the shape of the density deviation is different from $|\phi(x)|^2$ as well. Nonetheless, if $\frac{1}{d}<<q_c$ (as it is shown in fig. \ref{fig:spec}), the variation of $q$ around $q_c$ in $P_q$ can be neglected and $\langle e|\rho(x,0)|e\rangle=(1-P_{q_c})\,|\phi(x)|^2$. 

\subsection{\label{sec:ExcPrinc} The main approximations}
 Along the lines of the TL model, we neglect the fact that the particles deep below the Fermi point can jump between states. Next, it can be shown that the commutator  $[\widetilde{\rho}_{\eta,q},\widetilde{\rho}_{\eta',q'}]$ acts upon deep particles only (if ${|q|,|q'|<<k_F}$ ). It permits one to replace the real commutators of the densities with their averages \cite{Haldane81}. When the dispersion relation is linear, the commutator depending on time $[\widetilde{\rho}_{\eta,q}(t),\widetilde{\rho}_{\eta',q'}]$ can also be replaced with its average with the same great accuracy, because the density time dependence is simple $\widetilde{\rho}_{\eta,q}(t)= \widetilde{\rho}_{\eta,q} \,e^{-i \eta q\,\widetilde{v}_F t}$. Here we make the assumption that this approximation is valid for a nonlinear dispersion
\begin{eqnarray}
\label{eq:Commutators}
[\widetilde{\rho}_{\eta,q}(t),\widetilde{\rho}_{\eta',q'}]\approx \langle[\widetilde{\rho}_{\eta,q}(t),\widetilde{\rho}_{\eta',q'}] \rangle ,
 \end{eqnarray}
where the averaged commutator is easy to calculate by using (\ref{easyrhoT})
\begin{eqnarray}
	\langle[\widetilde{\rho}_{\eta,q}(t),\widetilde{\rho}_{\eta',q'}] \rangle =&& \frac{L}{2 \pi} \delta_{\eta,\eta'}\delta_{q,-q'} \eta q e^{-i \eta q \widetilde{v}_F t}\nonumber\\
	&&\times \frac{2 m^*}{q^2 t}\,\sin(\frac{q^2 t}{2 m^*}).
\end{eqnarray}

 If $m^*\rightarrow \infty$, this relation, as it should do, reduces to the well-known \cite{Haldane81} one of the case of a linear spectrum (up to the discussed time factor).  
 
 The approximation (\ref{eq:Commutators}) is not necessary to make, but in this case, the calculations and the final formulas become much more complicated making the effects we discuss here less clear. Anyway, it can be shown that the formulas coincide with each other when $d>>\lambda_F$, $d>>a_0$ and $w_0<<1$, so one may consider these relations as the conditions of applicability of (\ref{eq:Commutators}). Physically, this approximation means that the injected particle can create only a weak, slow varying field and, thus, slightly change the state of the liquid.
 
 \subsection{The calculation of the density evolution $\langle\rho_{q}(t)\rangle_e$ in the general quantum case}
Let us now obtain the main relations all the following is based on. By virtue of the relation (\ref{BogTran}), the calculation of (\ref{toCalculate}) comes down to the calculation of an object that contains the composite fermions density $\widetilde{\rho}_{\eta,q}(t)$, which time dependence (\ref{easyrhoT}) is known:
\begin{eqnarray}
	\label{ToCalculate}
	<\Psi_\eta(x_1) \widetilde{\rho}_{\eta',q}(t) \Psi_\eta^+(x_2)>
	=&&<\widetilde{\Psi}_\eta(x_1)\,\widetilde{F}_\eta(x_1) \widetilde{\rho}_{\eta',q}(t) \nonumber\\
	&&\times \widetilde{F}^+_\eta(x_2)\,\widetilde{\Psi}^+_\eta(x_2) >.
\end{eqnarray}

Throughout the calculations, we will mainly follow the method introduced in \cite{Rozhkov05} by Rozhkov. The main idea is that when $q \cdot \eta>0$, the operators $\tilde\rho_{\eta,q}$ act like annihilation operators  $\tilde\rho_{\eta,q}|0>=0$ and when  $q \cdot \eta<0$ like creation operators $<0|\tilde\rho_{\eta,q}=0$. The normal ordering, thereby, simplifies the construction (\ref{ToCalculate}). This calculation implies using the density commutators, and that is where the approximation (\ref{eq:Commutators}) is used. The calculation details are given in Appendix \ref{MainDeriv}, and the result is
\begin{widetext}
	\begin{subequations}
		\label{RHO_Wide}
		\begin{eqnarray}
			\label{RHO_r_Wide}
			\langle\widetilde\rho_{+,q}(t)\rangle_e=&&\frac{i}{2 \pi}\int(\frac{\alpha^2}{\alpha^2+y^2})^{u_0^2}\frac{\varphi^*_{q_1}\varphi_{q_1+q}e^{-i q_1y}}{2\pi(y-\frac{q t}{m^*}+i\alpha)}
			 e^{-i t(q\widetilde{v}_F+\frac{q^2}{2 m^*} )} d y d q_1 \nonumber\\
			&&+w_0 \int (1-P_{q_1})\frac{\varphi^*_{q_1}\varphi_{q_1+q}}{2\pi}
			  e^{-i t q\widetilde{v}_F }  \frac{2 m^*}{q^2 t}\,sin(\frac{q^2 t}{2 m^*})  d q_1 ,
		\end{eqnarray}
		\begin{eqnarray}
			\label{RHO_l_Wide}
			\langle\widetilde\rho_{-,q}(t)\rangle_e= u_0
			\int(1-P_{q_1})\frac{\varphi^*_{q_1}\varphi_{q_1+q}}{2\pi} 
			\times e^{i t q\widetilde{v}_F } \frac{2 m^*}{q^2 t}\,sin(\frac{q^2 t}{2 m^*})  d q_1 
		\end{eqnarray}
	\end{subequations}
\end{widetext}
for $q\geq0$. The formula of $q\leq0$ can be yielded by using $\langle\rho_{\eta,q}(t)\rangle^*_e=\langle\rho_{\eta,-q}(t)\rangle_e$. Here $\varphi_q=\phi_{k_F+q}$, which is more convenient. 

A complicated factor in (\ref{RHO_Wide}) is denoted as $1-P_q$, where $P_q$ is, in fact, the distribution function
\begin{eqnarray}
	\label{eq:equilibfunc}
	P_q=&&\langle 0| a^+_{k_F+q} a_{k_F+q}|0\rangle= \langle 0| c^+_{q} c_{q}|0\rangle\nonumber\\
	=&&1-\frac{i}{2 \pi}\int(\frac{\alpha^2}{\alpha^2+y^2})^{u_0^2}\frac{e^{-i q y}}{y+i\alpha}dy,
\end{eqnarray} 
that can be easily obtained from the well-known Green function \cite{Haldane81,Gogolin2004}.

We will see below that the two terms in (\ref{RHO_r_Wide}) describe the fast- and the slow-moving humps respectively and after applying (\ref{BogTran}), one obtains three humps, which is consistent with the numerical result in \cite{Moreno13}, except for, naturally, there is no spin fraction.

\subsection{The density evolution $\langle\rho_{q}(t)\rangle_e$ in known particular cases}
Let us now show that the result (\ref{RHO_Wide}) includes the formulas for different particular cases.

For $t=0$, the relations (\ref{RHO_Wide}) and (\ref{eq:rhoProport}) give the total initial density deviation of electrons
\begin{eqnarray}
	\label{eq:RhoT0}
	\langle\rho_{q}(0)\rangle_e=&&\int(1-P_{q_1})\frac{\varphi^*_{q_1}\varphi_{q_1+q}}{2\pi} dq_1\nonumber\\
	\approx&&(1-P_{q_c})\rho^0_q,
\end{eqnarray} 
where the approximation $q_c>>1/d$ is made, so that $P_q\approx P_{q_c}$.
The physical meaning of (\ref{eq:RhoT0}) was discussed in section \ref{sec:ExcPrinc}.

If $t \rightarrow 0$ or $m \rightarrow \infty$, the relations (\ref{RHO_Wide}) naturally reduce to the ones of the linear spectrum. For simplicity, suppose that $q_c>>\frac{1}{d}$. In the end, one obtains the relation (\ref{RHO_Lin}) up to the irrelevant factor $1-P_{q_c}$.

Next, it is shown in Appendix \ref{AppSpreading} that in the absence of interaction $u_0=0$; $w_0=0$, the relation (\ref{RHO_r_Wide}) can be reduced to the single free particle solution. The only difference here is that the Exclusion Principle remains in effect, and the distribution function $P_q$ (fig. $\ref{fig:spec}$) transforms into a step function. The density $\rho(x,t)$ dependence on $x$ of a free particle is shown in fig. \ref{fig:free}.

\subsection{The density evolution $\langle\rho_{q}(t)\rangle_e$ for intermediate times}
The result (\ref{RHO_Wide}) is quite general, but in some cases it will be more convenient for us to use its simplified version. Specifically, when time $t$ is presumed to be much lesser than the free spreading time $t<<m d^2$, but still beyond that of the TL theory. We also presume that the momentum distribution is narrow $q_c>>\frac{1}{d}$ (fig. \ref{fig:spec}). The relations (\ref{RHO_Wide}), then, immensely simplify to
\begin{subequations}
	\label{RHO_narrow}
	\begin{eqnarray}
		\langle\widetilde\rho_{+,q}(t)\rangle_e=&&\frac{i}{2 \pi}\int(\frac{\alpha^2}{\alpha^2+y^2})^{u_0^2}\frac{\rho^0_q e^{-i q_c y}}{y-\frac{q t}{m^*}+i\alpha} e^{-i t q\widetilde{v}_F } d y\nonumber\\
		&&+(1-P_{q_c})
		\rho^0_q  e^{-i t q\widetilde{v}_F } w_0  ,
		\label{RHO_narrow_right}
	\end{eqnarray}
	\begin{eqnarray}
		\langle\widetilde\rho_{-,q}(t)\rangle_e=(1-P_{q_c})
		\rho^0_q  e^{i t q\widetilde{v}_F } u_0   ,
		\label{RHO_narrow_left}
	\end{eqnarray}
\end{subequations}
where ``free'' initial density is defined as $\rho^0(x)=|\phi(x)|^2$ (see section \ref{sec:ExcPrinc}), so
\begin{equation}
	\label{eq:relDensPhi}
	\rho^0_q=\int\frac{\varphi^*_{q_1}\varphi_{q_1+q}}{2\pi} dq_1.
\end{equation}

\subsection{The slow humps}
Let us now discuss different characteristics of the humps; they can be analytically obtained from (\ref{RHO_Wide}), as it is shown in Appendix \ref{AppDeriv}. The slow humps move at $\pm\widetilde{v}_F$ (Appendix \ref{AppVel}) like they do in the semiclassical case (\ref{eq:VlasovSOL}). The time factor  $\frac{2 m^*}{q^2 t}\,sin(\frac{q^2 t}{2 m^*})$ means that the slow humps demonstrate a spreading $\propto \sqrt{t}$ (Appendix \ref{AppSpreading}) although we saw in the semiclassical case (\ref{eq:VlasovSOL}) that they do not spread at zero temperature. This effect is, thereby, quantum. For small times, when the humps have not spread yet, their amplitudes can be defined (Appendix \ref{AppAmpl}). The one of the left-moving slow hump remains the same \cite{Pham00,Steinberg08,Deshpande10, Das11} as in the case of a linear spectrum  $(1-K)/2$ , for the right-moving slow hump it becomes $(1+K)/2-\sqrt{K}$. Since the integral of the density does not depend on time, the amplitudes of the humps are proportional to $1/\sqrt{t}$ when time is large enough.   All the characteristic are summed up in table \ref{tab:CharsQuantum}.

 The slow humps differ in nature from the fast one, since they are plasmons of the liquid created at the beginning by the injected particle field. Indeed, any nonuniform density deviation creates an effective field that acts upon the liquid. Next it is easy to show that these plasmons can be as well excited by some external field $U(x,t)$ (see for a semiclassical example Appendix \ref{sec:SimEx}). The problem of the density evolution in the case of external field comes down to calculating the structure factor and has been solved earlier in a quite general case \cite{Imambekov12}. To reduce the solution obtained there to our case, one should take $q\rightarrow 0$ and $a_0 \rightarrow 0$ that gives the same  time factor  $\frac{2 m^*}{q^2 t}\,sin(\frac{q^2 t}{2 m^*})$ as in (\ref{RHO_Wide}).

 To better understand the physics of this quantum spreading, let us discuss the scattering from the weak field the injected particle creates. In classical one-dimensional case, the particles of the liquid just pass through this field and in the end have the same momentum as they did before. In quantum mechanics the scattering from a weak barrier takes place, and it changes the dynamics considerably. Similarly, field that starts acting upon the liquid at some moment leads to different further dynamic in classical and quantum cases.
 
  One more thing to notice here is that if the temperature was not zero, the slow humps would have an additional mechanism of spreading that would be linear in time, which is consistent with what we mentioned in the semiclassical consideration.

 \begin{figure}[]
\centerline{\includegraphics[width=1\linewidth]{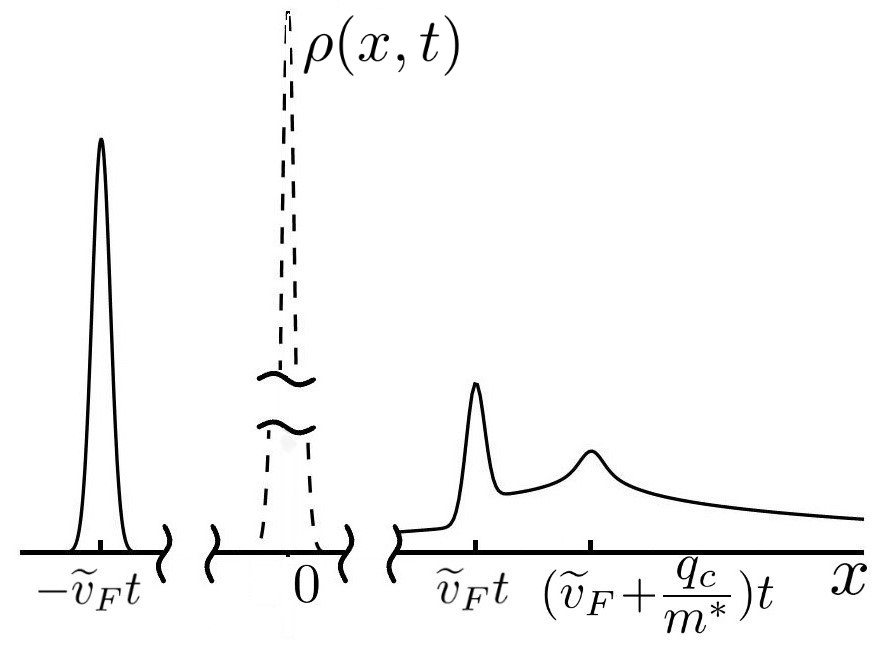}}
\caption{The density distributions when the interaction is strong enough; obtained by calculating (\ref{RHO_narrow}); time is not too large $t<<m d^2$, and the single-particle momentum distribution is narrow $q_c>>\frac{1}{d}$}
\label{fig:inter}
\end{figure}

\subsection{The fast hump}
The fast hump moves at $ \widetilde v_0=\widetilde{v}_F+q_c/m^*$ (Appendix \ref{AppVel}), unlike what it does in the semiclassical limit (\ref{eq:VlasovSOL}), where the velocity is $v_0=v_F+q_c/m$. For small times, the amplitude of the fast hump is $\sqrt{K}$. Note that when the dispersion relation is linear or time is small enough, the right-moving slow hump and the fast hump comove and add up to the usual hump with the known \cite{Pham00,Steinberg08,Deshpande10, Das11} amplitude $(1+K)/2$.  As for the slow humps, these characteristic are given in table \ref{tab:CharsQuantum}.

There are, broadly speaking, two mechanisms of changing the fast hump shape. The first mechanism is the discussed behavior a free particle wave packet demonstrates (fig. \ref{fig:free}), the fast-moving ``semiclassical'' hump (\ref{eq:VlasovSOL}) spreads in this way as well.  This ``free'' mechanism does not require any interaction, but in the presence of interaction, another mechanism occurs (fig. \ref{fig:inter}), and the stronger the interaction becomes, the more significant this effect is. We call it ``the spreading due to interaction'', and it has quantum nature, so it did not take place in the semiclassical case (\ref{eq:VlasovSOL}).

To consider the spreading due to interaction only, suppose the momentum distribution narrow $q_c>>\frac{1}{d}$, time $t$ small enough $t<<m d^2$ and the interaction strong enough, then the ``free'' spreading of the fast hump vanishes as well as the mechanism of spreading for the slow humps, since its characteristic time is the same $m d^2$. The relations (\ref{RHO_Wide}) simplify to (\ref{RHO_narrow}). 

The density distribution in real $x$ space is shown in fig. \ref{fig:inter} for a typical case. The most noticeable feature here is the long tail of the fast hump. Its asymptotic behavior in real space $x$ can be obtained from (\ref{RHO_narrow}) and is	
\begin{eqnarray}
	\label{eq:Tail}
	\langle\widetilde\rho_{+}(x,t)\rangle_e\rightarrow \frac{2 \pi u_0^2}{x-\widetilde v_0 t}
\end{eqnarray}
in the limit of large times $t>>\frac{m d}{q_c}$ (but still $t<<m d^2$) and $d<<x-\widetilde v_0 t<< v_F t$. 

The upper restriction $x-\widetilde v_0 t<< v_F t$ demonstrates a limitation of the theory. The tail (\ref{eq:Tail}) has, of course, to be integrable and, formally, the theory gives the factor $\exp (-\frac{4 \pi(x-\widetilde v_0 t)}{ v_F t})$ in (\ref{eq:Tail}). But it is just an adverse consequence of introducing the ``effective band-width'' parameter $\alpha$ \cite{Giamarchi03,vonDelft98} for a point-like interaction. If the interaction radius is finite, such issues do not occur \cite{Giamarchi03}.

The formula (\ref{eq:Tail}) shows that not only does the fast hump widen, but may also change its shape completely. In the general case, it demonstrates the free spreading $\Delta x_{free}(t)\sim\frac{1}{m d} t$. Additionally, the long tail (\ref{eq:Tail}) occurs because of the spreading due to interaction. The amplitude of the tail is proportional to the interaction strength square $u_0^2$, and the amplitude of the core part decreases with the interaction, so that the density integral is constant.

As for the physical meaning of the spreading due to interaction, we cannot provide any clear picture, just state that it relates to the mutual quantum scattering between the injected particle and the particles of the liquid, and that no detail here can be dropped or simplified. Anyway, this mechanism has a simple description in terms of the composite fermions, since they are free (\ref{eq:free}) and, thus, spread only in the ``free'' way. Injecting an electron is equivalent to excitation of a complicated superposition of the composite fermions, but their density is proportional to the density of the electrons (\ref{eq:rhoProport}), so all we need to do is analyze the non-equilibrium momentum distribution of the composite fermions. It can be shown that it is concentrated around $q=0$ and $q=q_c$, which corresponds to the slow and the fast humps. In addition to the initial momentum distribution of the injected electron, the part concentrated around $q=q_c$ has long tails that are proportional to the derivative of the equilibrium distribution function of the electrons $P'_{q_c-q}$ (Appendix \ref{AppSpreading}). These tails result in the addition spreading of the composite fermions density, which is the spreading due to interaction.

\subsection{The limits of applicability}
We have completely neglected the interaction between the composite fermions that may change the dynamics at large time scales. Specifically, we considered neither cubic nor quadratic terms in (\ref{HamiKvasi}), however, for both of these terms, the smaller the radius of the interaction is, the larger the time of applicability of the free approximation can be. Also, we did not consider the ``point-like'' interaction between quasiparticles on opposite branches, but the time of applicability increases with the characteristic size $d$ of the density deviation.

The next restriction is that the mentioned gradient catastrophe time has to be much larger than the characteristic time of spreading $t_{grad}>>md^2$. In our process this time can be estimated as $t_{grad}=\sqrt{m d^3/(g_0 N)}$ (Appendix \ref{sec:tgrad}) and gives a macroscopic time if the number of the injected particles $N=1$ and the other parameters are typical. This estimate differs from that of the case in \cite{Bettelheim12}, where a system has time to relax to a local equilibrium $t_{grad}=\frac{m d}{\Delta \rho}$, where $\Delta \rho$ is the amplitude of the density deviation.

\begin{table}[]
	\caption{\label{tab:CharsQuantum}
		The humps characteristics obtained by analysis of the quantum solution (\ref{RHO_Wide}). The amplitudes are yielded in the limit of small time. As usual, they are scaled so that their sum always equals $1$.
	}
	\begin{ruledtabular}
		\begin{tabular}{c c c c}
			              Hump               & Velocity                 & Amplitude                & Spreading                                                                                                    \\ \hline\hline
			              Fast               & $\widetilde v_F+q_c/m^*$ & $\sqrt{K}$               & \shortstack{Free spreading\footnotemark[1] $\propto t$ \\and\\ spr.\footnotemark[2] due to int. $\propto t$} \\ \hline
			\shortstack{Right-moving \\slow} & $\widetilde v_F$         & $\frac{1+K}{2}-\sqrt{K}$ & Spreading\footnotemark[3] $\propto\sqrt{t}$                                                                 \\ \hline
			\shortstack{Left-moving \\slow}  & $-\widetilde v_F$        & $\frac{1-K}{2}$          & Spreading\footnotemark[3] $\propto\sqrt{t}$
		\end{tabular}
	\end{ruledtabular}
\footnotetext[1]{The simple spreading behavior a free particle wave packet demonstrates (fig. \ref{fig:free})}
\footnotetext[2]{The quantum spreading due to interaction that relates to the mutual quantum scattering between the injected particle and the particles of the liquid}
\footnotetext[3]{The spreading proportional to $\propto\sqrt{t}$ associated with quantum scattering of the particles of the liquid from the field the injected particle creates.}
\end{table}

\section{Conclusions}

We have analytically considered the density evolution of a spinless Fermi liquid with a nonlinear dispersion relation into which one particle is injected. The interaction is short-ranged and the temperature is zero; the electron momentum distribution function is concentrated near the right Fermi point. In the case of a linear dispersion relation, this problem was studied long ago by means of TL model \cite{Das11}. However, this linear model is valid only for small times, and to describe the fractionalization effect over longer timescales, it is necessary to take the nonlinearity of the spectrum into account.

 The current state of the theory has allowed us to examine the problem analytically in the case of a nonlinear spectrum. It is known that for a linear spectrum, two opposite moving humps of the density occur, and they simply move at $\pm \widetilde v_F$ without changing their shapes. In the nonlinear liquid, there are generally three humps of the density and they change their shapes in a complicated manner. 
 
  Two of the humps are the plasmons of the liquid moving at about $+ \widetilde v_F$ or $- \widetilde v_F$, we call them ``slow humps''. They are excited by the initial hump field at the beginning of the movement. In the limit of small time, the amplitudes of the left- and right-moving ones are $(1-K)/2$ and $(1+K)/2-\sqrt{K}$ respectively. The slow humps spread proportionally to $\sqrt{t}$. This spreading has quantum nature and relates to the scattering from a weak barrier.

  Another hump is fast and moves at $\widetilde{v}_F+q_c/m^*$ to the right, it corresponds to the injected particle. Its amplitude is $\sqrt{K}$ for small times. Apart from the simple free mechanism, the fast hump has a quantum mechanism of spreading due to interaction, which is associated with the mutual quantum scattering between the injected particle and the particles of the liquid.

 We also showed that the fractionalization into three humps as well as the free mechanism of spreading can be illustrated by the classical Vlasov equation, where the equilibrium distribution function is quantum-like. Moreover, in the case of a linear dispersion relation, this solution is the same as the well-known TL model solution.

\begin{acknowledgments}
We thank I. V. Gorny for valuable discussions. We also thank D.G. Polyakov for the conversation about the results. This work was supported by the Russian Science Foundation (grant No. 18-02-01016-a).
\end{acknowledgments}

\appendix
\begin{widetext}
\section{\label{MainDeriv}}

Normal ordering means shifting the density operators to the right or left part, correspondingly. It allows one to recast (\ref{ToCalculate}) to the form depending on free operators only:
\begin{subequations}
\begin{eqnarray*}
\langle\Psi_+(x_1) \widetilde{\rho}_{+,q}(t) \Psi_+^+(x_2)\rangle=&&(\frac{\alpha^2}{\alpha^2+(x_1-x_2)^2})^{\sinh^2(\theta_0)}\langle\widetilde{\Psi}_+(x_1) \widetilde{\rho}_{+,q}(t)\widetilde{\Psi}_+^+(x_2)\rangle\\
&&+w_0(\frac{\alpha^2}{\alpha^2+(x_1-x_2)^2})^{\sinh^2(\theta_0)}\langle \widetilde{\Psi}_+(x_1) \widetilde{\Psi}_+^+(x_2) \rangle  e^{-i q \widetilde v_F t}\frac{2 m^*}{q^2 t}\,sin(\frac{q^2 t}{2 m^*}) [\theta(q)e^{-i q x_2}+\theta(-q)e^{-i q x_1}],
\end{eqnarray*}
\begin{eqnarray*}
\langle\Psi_+(x_1) \widetilde{\rho}_{-,q}(t) \Psi_+^+(x_2)\rangle=&&(\frac{\alpha^2}{\alpha^2+(x_1-x_2)^2})^{\sinh^2(\theta_0)}\langle\widetilde{\Psi}_+(x_1) \widetilde{\rho}_{-,q}(t)\widetilde{\Psi}_+^+(x_2)\rangle\\
&&+u_0(\frac{\alpha^2}{\alpha^2+(x_1-x_2)^2})^{\sinh^2(\theta_0)}\langle \widetilde{\Psi}_+(x_1) \widetilde{\Psi}_+^+(x_2) \rangle  e^{i q \widetilde v_F t}\frac{2 m^*}{q^2 t}\,sin(\frac{q^2 t}{2 m^*}) [\theta(q)e^{-i q x_2}+\theta(-q)e^{-i q x_1}].
\end{eqnarray*}
\end{subequations}
After using Wick's theorem one obtains
\begin{subequations}
\begin{eqnarray*}
\int e^{i q_2 x_2-iq_1 x_1}\langle\Psi_+(x_1) \widetilde{\rho}_{+,q}(t) \Psi_+^+(x_2)\rangle dx_1 dx_2
=&&\int(\frac{\alpha^2}{\alpha^2+y^2})^{\sinh^2(\theta_0)} \frac{\delta(q_2-q_1-q)}{(y-\frac{q t}{m^*}+i\alpha)} i e^{-i q\widetilde{v}_F t}\\
&&\times(\theta(q)e^{i(-q_1 y-\frac{q^2 t}{2 m^*})}+\theta(-q)e^{i(-q_2 y+\frac{q^2 t}{2 m^*})})dy\\
&&+\int(\frac{\alpha^2}{\alpha^2+y^2})^{\sinh^2(\theta_0)}\frac{\delta(q_2-q_1-q)}{ (y+i \alpha)}i w_0 e^{-i q\widetilde{v}_F t}\frac{2 m^*}{q^2 t}\,sin(\frac{q^2 t}{2 m^*})\\
&&\times [\theta(q)e^{-i q_1 y}+\theta(-q)e^{-i q_2 y}]dy
\end{eqnarray*}
\begin{eqnarray*}
\int e^{i q_2 x_2-iq_1 x_1}\langle\Psi_+(x_1) \widetilde{\rho}_{-,q}(t) \Psi_+^+(x_2)\rangle dx_1 dx_2=&&
\int(\frac{\alpha^2}{\alpha^2+y^2})^{\sinh^2(\theta_0)}\frac{\delta(q_2-q_1-q)}{ (y+i \alpha)}i u_0 e^{i q\widetilde{v}_F t}\frac{2 m^*}{q^2 t}\,sin(\frac{q^2 t}{2 m^*})\\
&&\times [\theta(q)e^{-i q_1 y}+\theta(-q)e^{-i q_2 y}]dy,
\end{eqnarray*}
\end{subequations}
where $y=x_1-x_2$.

To yield the density behavior $\langle\rho_{\eta,q}(t)\rangle_e$ of a real particle that has the momentum distribution $\phi_{q}$, the relation $\int e^{i q_2 x_2-iq_1 x_1}<\Psi_\eta(x_1) \rho_{\eta',q}(t)\Psi_\eta^+(x_2)>$ should be multiplied by $\frac{\phi_{q_1}^* \phi_{q_2}}{(2\pi)^2}$ and then integrated over $q_1$ and $q_2$, which leads to (\ref{RHO_Wide})
\newpage
\end{widetext}

\section{\label{AppDeriv} The derivation of the humps characteristics}
\subsection{\label{AppAmpl}The amplitudes}

Let us now obtain the amplitudes of the humps in the limit of small time t$\rightarrow$0, so we will use (\ref{RHO_narrow}). As one can see from (\ref{RHO_narrow_left}), for the left-moving slow hump, its amplitude in terms of composite fermions is $u_0$, where the irrelevant factor $1-P_{q_c}$ is dropped.   For the electrons, one can derive from (\ref{eq:rhoProport}) that the density amplitude is $u_0 \sqrt{K}=\frac{1}{2}(1/\sqrt{K}-\sqrt{K})\sqrt{K}=(1-K)/2 $. Similarly, for the right-moving slow hump, the composite fermions density amplitude is $w_0$, and it equals to $1$ for the fast one, which gives us the results shown in table \ref{tab:CharsQuantum}.
\subsection{\label{AppVel}The velocities}
One does not have to consider the spreading when deriving the hump velocities, so it is easier to use (\ref{RHO_narrow}) for this purpose. For the left-moving slow hump, the only factor that depends on q (apart from $\rho^0_q$) is $e^{i t q\widetilde{v}_F }$, so, similarly to (\ref{eq:FurierTransfShift}), one obtains 
\begin{eqnarray}
\langle \widetilde \rho(x,t)\rangle=&&(1-P_{q_c})\,u_0  \int \frac{dq}{2 \pi} 	\rho^0_q  e^{i  q(x+ t\widetilde{v}_F) }\nonumber\\
=&&(1-P_{q_c})\,u_0 \widetilde \rho^0(x+ t\widetilde{v}_F),
\end{eqnarray}
 and the velocity is $-\widetilde{v}_F$. Similarly, for the right-moving slow hump it is $\widetilde{v}_F$.

It is easy to find the fast hump velocity if the interaction is zero, since the calculation of (\ref{RHO_narrow_right}) is reduced to an integration around a small circle about the pole $y=\frac{q t}{m}-i \alpha$; the ``free'' velocity is $q_c/m+v_F$. In the presence of interaction, in addition to the integration around the pole, one has to calculate an integral along the boundary between the sheets of the Riemann surface described by the factor $(\frac{\alpha^2}{\alpha^2+y^2})^{u_0^2}$; however, if the interaction is weak and $t$ is large enough($t>>\frac{m d}{q_c}$, but still $t<<m d^2$), the latter integral can be neglected, and the first term in (\ref{RHO_narrow_right}) approximately equals to
\begin{eqnarray}
	\label{eq:PoleInt}
	\langle\widetilde\rho^{(fs)}_{+,q}(t)\rangle_e=&&\frac{i}{2 \pi}\int(\frac{\alpha^2}{\alpha^2+y^2})^{u_0^2}\frac{\rho^0_q e^{-i q_c y}}{y-\frac{q t}{m^*}+i\alpha} e^{-i t q\widetilde{v}_F } d y\nonumber\\
	\approx&& (\frac{\alpha^2}{\alpha^2+(\frac{q t}{m^*}-i\alpha)^2})^{u_0^2}\rho^0_qe^{-i q_c (\frac{q t}{m^*}-i\alpha)} e^{-i t q\widetilde{v}_F }. \nonumber\\
\end{eqnarray}
Now the factor $(...)^{u_0^2}$ reflects the spreading due to interaction, and, as for the slow humps, it is easy to find the velocity, which is $\widetilde{v}_F+q_c/m^*$. These results are summed up in table \ref{tab:CharsQuantum}.

\subsection{\label{AppSpreading} The spreading}
 Let us start with the slow humps spreading, which can be obtained from (\ref{RHO_Wide}). 
 It is described by the factor $\frac{2 m^*}{q^2 t}\,sin(\frac{q^2 t}{2 m^*})$ and when $t$ is large enough ($t>>m d^2$), the simple estimate for the characteristic width $\Delta q$ in momentum space $\frac{(\Delta q)^2 t}{m^*}\approx1$ gives the result ${\Delta x(t)\sim 1/\Delta q\approx\sqrt{t/m^*}}$.

Let us now show  that the result (\ref{RHO_r_Wide}) includes the free spreading mechanism; more specifically, that it is the only one mechanism in the absence of interaction $u_0=0$ and $w_0=0$. To avoid the exclusion principle influence discussed in section \ref{sec:ExcPrinc}, assume that $\varphi(q_1)=0$ if $q_1<0$, then, after integration around the pole $y=\frac{q t}{m}-i \alpha$ and taking the inverse Fourier transform, the relation (\ref{RHO_r_Wide}) reduces to 
\begin{eqnarray}
	\label{eq:freemech}
	\langle\rho^{(free)}_{+}(x,t)\rangle_e=&&\frac{1}{(2\pi)^2}\int\varphi^*(q_1)\varphi(q_1+q)e^{-i q_1 (\frac{q t}{m^*}-i\alpha)}\nonumber\\
	&&\times e^{-i t(q v_F+\frac{q^2}{2 m^*} )+i q x} d q_1 dq,
\end{eqnarray}
where we must assume that $\alpha q_1<<1$. Next, the free particle wave function is
\[
\phi(x,t)=\frac{1}{2\pi}\int dq_1 \varphi(q_1) e^{i [(q_1+k_F) x-\frac{t (q_1+k_F)^2 }{2 m^*}]},
\]
and the density $\langle\rho(x,t)\rangle_e=\phi(x,t) \phi^*(x,t)$ after variable replacement becomes (\ref{eq:freemech}). 

The last mechanism to consider is spreading due to interaction. Let us show that it is equivalent to spreading that is linear on time with an effective distribution that depends on the derivative $P'_q$ of the equilibrium distribution.  The equilibrium distribution function $P_q$ is written in the form (\ref{eq:equilibfunc}). Its derivative is 
\begin{eqnarray}
	\label{eq:equilibfuncDeriv}
	P'_q=&&\frac{1}{2 \pi}\int(\frac{\alpha^2}{\alpha^2+y^2})^{u_0^2}\frac{ y e^{-i q y}}{y+i\alpha}dy\nonumber\\
	\approx&&\frac{1}{2 \pi}\int(\frac{\alpha^2}{\alpha^2+y^2})^{u_0^2} e^{-i q y}dy,
\end{eqnarray}
where we used $q \alpha<<1$, so only large $y$ contributes to the integral. Rewrite the relation (\ref{RHO_narrow_right}) to see that it contains the derivative (\ref{eq:equilibfuncDeriv}). Start with the mentioned simplification (\ref{eq:PoleInt}), it can also be rewritten as 
\begin{widetext}
\begin{eqnarray}
	\langle\widetilde\rho^{(fs)}_{+,q}(t)\rangle_e
	\approx&& (\frac{\alpha^2}{\alpha^2+(\frac{q t}{m^*}-i\alpha)^2})^{u_0^2}\rho^0_qe^{-i q_c (\frac{q t}{m^*}-i\alpha)} e^{-i t q\widetilde{v}_F }	
	=\int dy(\frac{\alpha^2}{\alpha^2+y^2})^{u_0^2}\rho^0_q e^{-i q_c y-i t q\widetilde{v}_F }\delta(y-\frac{q t}{m^*}+i\alpha)\nonumber\\
	=&&\int dy(\frac{\alpha^2}{\alpha^2+y^2})^{u_0^2}\rho^0_q e^{-i q_c y-i t q\widetilde{v}_F }\int \frac{dw}{2\pi}e^{i w (y-\frac{q t}{m^*}+i\alpha)}
	=\int dw\, P'_{q_c-w}\rho^0_q e^{i w (-\frac{q t}{m^*})-i t q\widetilde{v}_F},	
\end{eqnarray}
which shows the role of the ground state momentum distribution function in the spreading due to interaction.
\end{widetext}

\begin{figure}[h]
	\centerline{\includegraphics[width=1\linewidth]{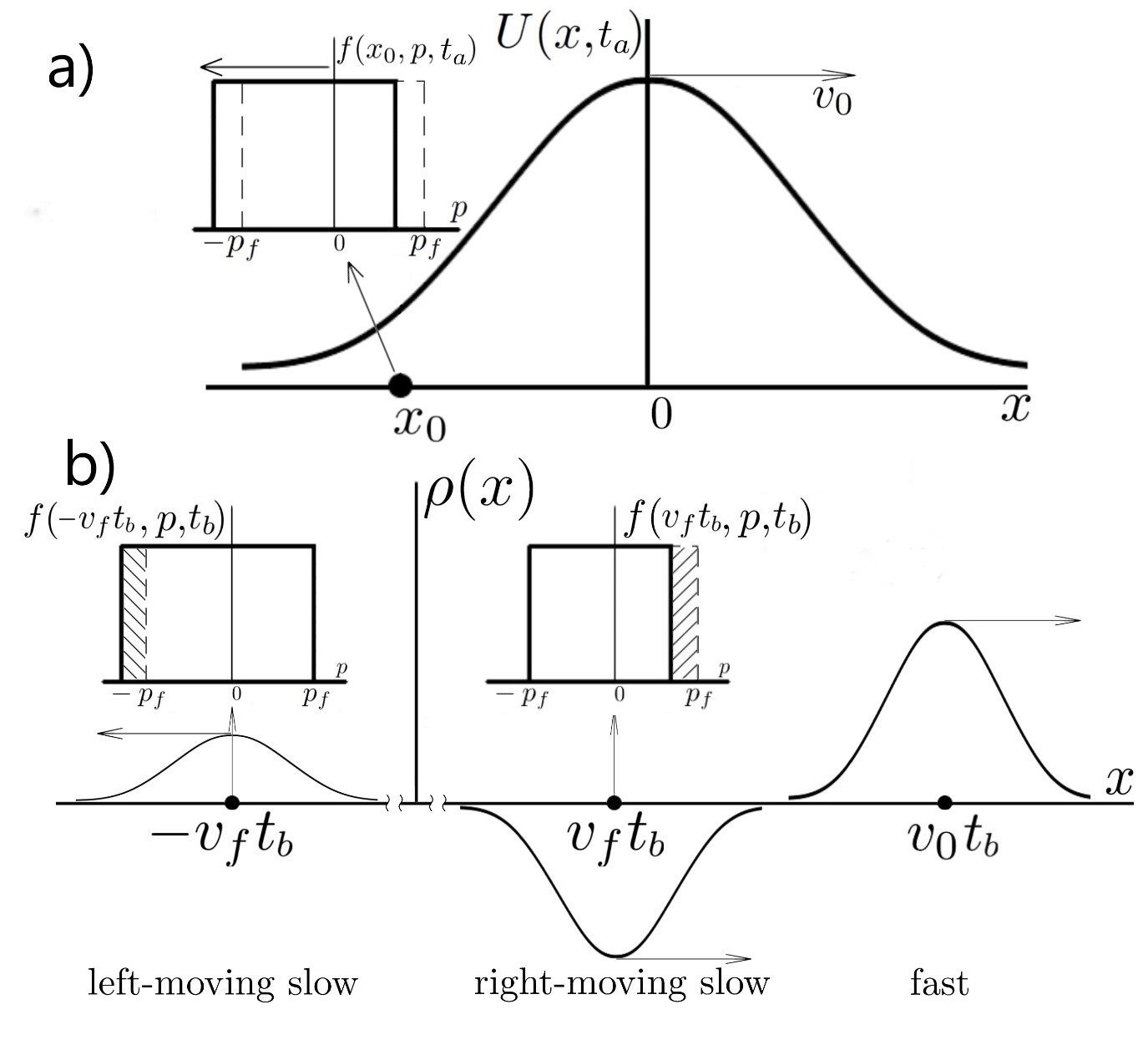}}
	\caption{The semiclassical problem of an external field applied to a free Fermi gas. In fig. a, it is shown the field that has just appeared and is moving to the right, so $t_a<<d/v_f$. The field causes shifting of all particles to the left in momentum space. Effectively, just the particles close to the Fermi points change their states. It is shown in the embedded picture of momentum distribution at some point $x_0$. Ultimately, there is an excess of left-moving particles and a lack of right-moving ones. They form the left- and right-moving slow humps respectively (fig. b) when the time is large enough, and all three humps split from each other $t_b > d/(v_0-v_f)$. There is also a density deviation comoving with the field, it corresponds to the fast hump. One can see that, for instance, the left-moving density deviation consists of the particles with the velocity $\approx v_f$.}
	\label{fig:SimEx}
\end{figure}

\subsection{ \label{sec:SimEx} A simple example of fractionalization}
 
To understand the fractionalization effect better, let us consider the simplest system we managed to find that demonstrates it. It is a classical system of free particles that has a step momentum distribution.
\begin{equation}
	f_0(p)=\theta(p_F-|p|).
\end{equation}

Instead of injecting a particle into the system, apply an external field $U(x,t)$ that appears at $t=0$ and moves to the right at constant velocity $v_0$, so that 
\begin{equation}
	\label{eq:AppExtFd}
	U(x,t)=\theta(t) U_0(x-v_0 t).
\end{equation}
Note that the injected particle from the semiclassical consideration (Section \ref{sec:SemiclConsid}) creates similar field, so the systems are fairly close. This model is classical, so the particles of the liquid just pass through the field (\ref{eq:AppExtFd}) and in the end have the same momentum as they did before the field started acting upon them. However, this is not the case for the particles that were within the field when it appeared, the field effect is uncompensated for them, and these particles form the slow humps eventually. The solution of (\ref{eq:Vlasov}) and explanations are shown in fig. \ref{fig:SimEx}.

The situation is more profound in the case of interacting liquid and an injected particle; for instance, the right-moving particles always comove with the left-moving ones (fig. \ref{fig:Lin}), but the main reason of fractionalization is the same.

\begin{figure}[h!]
	\centerline{\includegraphics[width=0.75\linewidth]{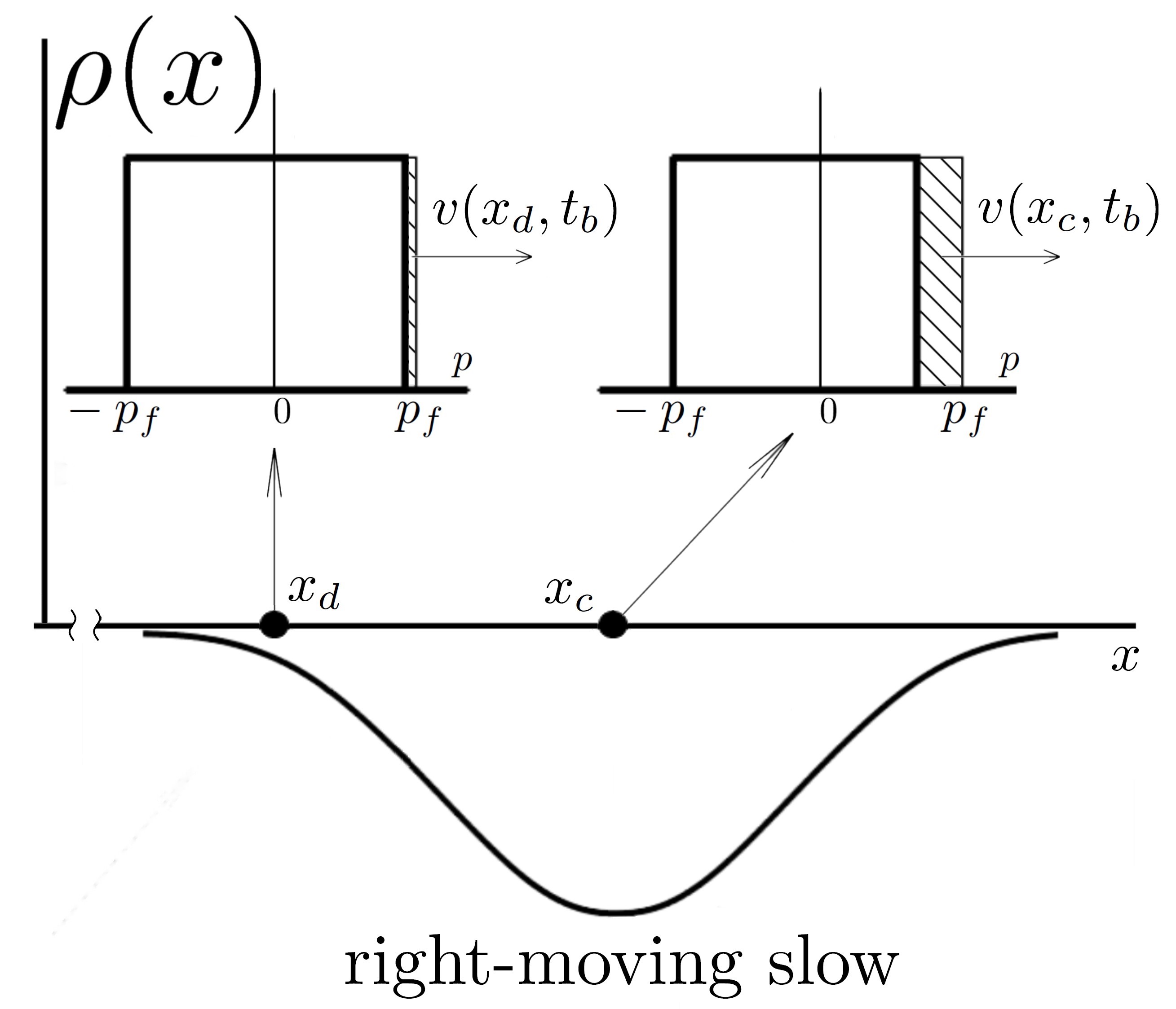}}
	\caption{It is shown the difference in averaged velocities of different areas of the right-moving slow density hump. The averaged velocity $v(x_c,t_b)$ of the top of the hump is slightly lower than its foot ${v(x_d,t_b)\rightarrow v_f}$, which leads to the gradient catastrophe when time is large enough $t\approx t_{grad}\approx d/\Delta v$, where $\Delta v=|v(x_c,t_b)-v(x_d,t_b)|$}
	\label{fig:SimEx_shock}
\end{figure}

\subsection{\label{sec:tgrad}The critical time of the gradient catastrophe}
Let us provide a simple lower estimate of the gradient catastrophe time, it is based on classical mechanics since this effect is classical. Generally speaking, the gradient catastrophe occurs if one region of a density deviation moves faster than another, and initially smooth front develops large gradients. Consequently, if $\Delta v$ is the maximum velocity deviation, then the critical formation time can be estimated as $t_{grad}\approx d/\Delta v$, where $d$ is the density deviation characteristic size. 

To find $\Delta v$ in our particular case, inject $N$ particles into a liquid, they create some moving field that acts upon the particles of the liquid and causes both the velocity change and the density deviation. For our estimate, we threat this filed as it is external and assume that it does not change its shape and amplitude. This means neglecting the backwards action upon the injected particles as well as their mutual repulse and spreading. We also neglect the interaction between the particles of the liquid. All these effects, however, can only weaker the field, which makes $\Delta v$ smaller and $t_{grad}$ larger, but we would just like to estimate the time from below. 

The problem, thus, comes down to the simple one considered in Appendix \ref{sec:SimEx}: a classical system of free particles that has step momentum distribution with external field (\ref{eq:AppExtFd}) acting upon it. The solution is shown in fig. \ref{fig:SimEx}, but some further clarification has to be made. The averaged velocities (local velocities) of different parts of the slow humps are not equal, since the field has acted on the constituent particles not equally (fig. \ref{fig:SimEx_shock}), the field affects the right-moving particles the most, so we focus on them.

Let us move to the coordinate system, where the external field is stationary. Once the field appears, some particles immediately come to be on a potential energy hump, it is they that form the slow humps in the end (Section \ref{sec:SimEx}). In this coordinate system, the velocity of the right-moving particles is $v_f-v_0$, which is presumed small. The particles increase their velocities while leaving the potential hump and do not change them afterwards. Note that in laboratory coordinate system, leaving the potential hump means that the slow humps, and the field split. 

The particles at the foot of the hump slightly change their initial velocity $v_f-v_0$. The particles at the top of the hump, obviously, achieve the maximum velocity $-v_b$, which is negative in this coordinate system. It can be found from the energy conservation law $m v_b^2/2=\Delta U$, where we neglect $v_f-v_0$. The amplitude of the potential energy hump is $\Delta U \approx g_0 N/d$. So $\Delta v=|v(x_c,t_b)-v(x_d,t_b)| \approx v_b$ and $t_{grad}\approx d/\Delta v\approx\sqrt{m d^3/(g_0 N)}$.

\bibliography{DD_20_biblio}

\begin{thebibliography}{31}%
\makeatletter
\providecommand \@ifxundefined [1]{%
 \@ifx{#1\undefined}
}%
\providecommand \@ifnum [1]{%
 \ifnum #1\expandafter \@firstoftwo
 \else \expandafter \@secondoftwo
 \fi
}%
\providecommand \@ifx [1]{%
 \ifx #1\expandafter \@firstoftwo
 \else \expandafter \@secondoftwo
 \fi
}%
\providecommand \natexlab [1]{#1}%
\providecommand \enquote  [1]{``#1''}%
\providecommand \bibnamefont  [1]{#1}%
\providecommand \bibfnamefont [1]{#1}%
\providecommand \citenamefont [1]{#1}%
\providecommand \href@noop [0]{\@secondoftwo}%
\providecommand \href [0]{\begingroup \@sanitize@url \@href}%
\providecommand \@href[1]{\@@startlink{#1}\@@href}%
\providecommand \@@href[1]{\endgroup#1\@@endlink}%
\providecommand \@sanitize@url [0]{\catcode `\\12\catcode `\$12\catcode
  `\&12\catcode `\#12\catcode `\^12\catcode `\_12\catcode `\%12\relax}%
\providecommand \@@startlink[1]{}%
\providecommand \@@endlink[0]{}%
\providecommand \url  [0]{\begingroup\@sanitize@url \@url }%
\providecommand \@url [1]{\endgroup\@href {#1}{\urlprefix }}%
\providecommand \urlprefix  [0]{URL }%
\providecommand \Eprint [0]{\href }%
\providecommand \doibase [0]{https://doi.org/}%
\providecommand \selectlanguage [0]{\@gobble}%
\providecommand \bibinfo  [0]{\@secondoftwo}%
\providecommand \bibfield  [0]{\@secondoftwo}%
\providecommand \translation [1]{[#1]}%
\providecommand \BibitemOpen [0]{}%
\providecommand \bibitemStop [0]{}%
\providecommand \bibitemNoStop [0]{.\EOS\space}%
\providecommand \EOS [0]{\spacefactor3000\relax}%
\providecommand \BibitemShut  [1]{\csname bibitem#1\endcsname}%
\let\auto@bib@innerbib\@empty
\bibitem [{\citenamefont {Bocquillon}\ \emph {et~al.}(2014)\citenamefont
  {Bocquillon}, \citenamefont {Freulon}, \citenamefont {Parmentier},
  \citenamefont {Berroir}, \citenamefont {Pla{\c{c}}ais}, \citenamefont {Wahl},
  \citenamefont {Rech}, \citenamefont {Jonckheere}, \citenamefont {Martin},
  \citenamefont {Grenier} \emph {et~al.}}]{Bocquillon14}%
  \BibitemOpen
  \bibfield  {author} {\bibinfo {author} {\bibfnamefont {E.}~\bibnamefont
  {Bocquillon}}, \bibinfo {author} {\bibfnamefont {V.}~\bibnamefont {Freulon}},
  \bibinfo {author} {\bibfnamefont {F.~D.}\ \bibnamefont {Parmentier}},
  \bibinfo {author} {\bibfnamefont {J.-M.}\ \bibnamefont {Berroir}}, \bibinfo
  {author} {\bibfnamefont {B.}~\bibnamefont {Pla{\c{c}}ais}}, \bibinfo {author}
  {\bibfnamefont {C.}~\bibnamefont {Wahl}}, \bibinfo {author} {\bibfnamefont
  {J.}~\bibnamefont {Rech}}, \bibinfo {author} {\bibfnamefont {T.}~\bibnamefont
  {Jonckheere}}, \bibinfo {author} {\bibfnamefont {T.}~\bibnamefont {Martin}},
  \bibinfo {author} {\bibfnamefont {C.}~\bibnamefont {Grenier}}, \emph
  {et~al.},\ }\bibfield  {title} {\bibinfo {title} {Electron quantum optics in
  ballistic chiral conductors},\ }\href@noop {} {\bibfield  {journal} {\bibinfo
   {journal} {Annalen der Physik}\ }\textbf {\bibinfo {volume} {526}},\
  \bibinfo {pages} {1} (\bibinfo {year} {2014})}\BibitemShut {NoStop}%
\bibitem [{\citenamefont {Grenier}\ \emph {et~al.}(2011)\citenamefont
  {Grenier}, \citenamefont {Herv{\'e}}, \citenamefont {F{\`e}ve},\ and\
  \citenamefont {Degiovanni}}]{Grenier11}%
  \BibitemOpen
  \bibfield  {author} {\bibinfo {author} {\bibfnamefont {C.}~\bibnamefont
  {Grenier}}, \bibinfo {author} {\bibfnamefont {R.}~\bibnamefont {Herv{\'e}}},
  \bibinfo {author} {\bibfnamefont {G.}~\bibnamefont {F{\`e}ve}},\ and\
  \bibinfo {author} {\bibfnamefont {P.}~\bibnamefont {Degiovanni}},\ }\bibfield
   {title} {\bibinfo {title} {Electron quantum optics in quantum hall edge
  channels},\ }\href@noop {} {\bibfield  {journal} {\bibinfo  {journal} {Modern
  Physics Letters B}\ }\textbf {\bibinfo {volume} {25}},\ \bibinfo {pages}
  {1053} (\bibinfo {year} {2011})}\BibitemShut {NoStop}%
\bibitem [{\citenamefont {B{\"a}uerle}\ \emph {et~al.}(2018)\citenamefont
  {B{\"a}uerle}, \citenamefont {Glattli}, \citenamefont {Meunier},
  \citenamefont {Portier}, \citenamefont {Roche}, \citenamefont {Roulleau},
  \citenamefont {Takada},\ and\ \citenamefont {Waintal}}]{Bauerle18}%
  \BibitemOpen
  \bibfield  {author} {\bibinfo {author} {\bibfnamefont {C.}~\bibnamefont
  {B{\"a}uerle}}, \bibinfo {author} {\bibfnamefont {D.~C.}\ \bibnamefont
  {Glattli}}, \bibinfo {author} {\bibfnamefont {T.}~\bibnamefont {Meunier}},
  \bibinfo {author} {\bibfnamefont {F.}~\bibnamefont {Portier}}, \bibinfo
  {author} {\bibfnamefont {P.}~\bibnamefont {Roche}}, \bibinfo {author}
  {\bibfnamefont {P.}~\bibnamefont {Roulleau}}, \bibinfo {author}
  {\bibfnamefont {S.}~\bibnamefont {Takada}},\ and\ \bibinfo {author}
  {\bibfnamefont {X.}~\bibnamefont {Waintal}},\ }\bibfield  {title} {\bibinfo
  {title} {Coherent control of single electrons: a review of current
  progress},\ }\href@noop {} {\bibfield  {journal} {\bibinfo  {journal}
  {Reports on Progress in Physics}\ }\textbf {\bibinfo {volume} {81}},\
  \bibinfo {pages} {056503} (\bibinfo {year} {2018})}\BibitemShut {NoStop}%
\bibitem [{\citenamefont {Das}\ and\ \citenamefont {Rao}(2011)}]{Das11}%
  \BibitemOpen
  \bibfield  {author} {\bibinfo {author} {\bibfnamefont {S.}~\bibnamefont
  {Das}}\ and\ \bibinfo {author} {\bibfnamefont {S.}~\bibnamefont {Rao}},\
  }\bibfield  {title} {\bibinfo {title} {Spin-polarized scanning-tunneling
  probe for helical luttinger liquids},\ }\href@noop {} {\bibfield  {journal}
  {\bibinfo  {journal} {Physical review letters}\ }\textbf {\bibinfo {volume}
  {106}},\ \bibinfo {pages} {236403} (\bibinfo {year} {2011})}\BibitemShut
  {NoStop}%
\bibitem [{\citenamefont {Deshpande}\ \emph {et~al.}(2010)\citenamefont
  {Deshpande}, \citenamefont {Bockrath}, \citenamefont {Glazman},\ and\
  \citenamefont {Yacoby}}]{Deshpande10}%
  \BibitemOpen
  \bibfield  {author} {\bibinfo {author} {\bibfnamefont {V.~V.}\ \bibnamefont
  {Deshpande}}, \bibinfo {author} {\bibfnamefont {M.}~\bibnamefont {Bockrath}},
  \bibinfo {author} {\bibfnamefont {L.~I.}\ \bibnamefont {Glazman}},\ and\
  \bibinfo {author} {\bibfnamefont {A.}~\bibnamefont {Yacoby}},\ }\bibfield
  {title} {\bibinfo {title} {Electron liquids and solids in one dimension},\
  }\href@noop {} {\bibfield  {journal} {\bibinfo  {journal} {Nature}\ }\textbf
  {\bibinfo {volume} {464}},\ \bibinfo {pages} {209} (\bibinfo {year}
  {2010})}\BibitemShut {NoStop}%
\bibitem [{\citenamefont {Haldane}(1981)}]{Haldane81}%
  \BibitemOpen
  \bibfield  {author} {\bibinfo {author} {\bibfnamefont {F.}~\bibnamefont
  {Haldane}},\ }\bibfield  {title} {\bibinfo {title} {Luttinger liquid theory
  of one-dimensional quantum fluids. i. properties of the luttinger model and
  their extension to the general 1d interacting spinless fermi gas},\
  }\href@noop {} {\bibfield  {journal} {\bibinfo  {journal} {Journal of Physics
  C: Solid State Physics}\ }\textbf {\bibinfo {volume} {14}},\ \bibinfo {pages}
  {2585} (\bibinfo {year} {1981})}\BibitemShut {NoStop}%
\bibitem [{\citenamefont {Pham}\ \emph {et~al.}(2000)\citenamefont {Pham},
  \citenamefont {Gabay},\ and\ \citenamefont {Lederer}}]{Pham00}%
  \BibitemOpen
  \bibfield  {author} {\bibinfo {author} {\bibfnamefont {K.-V.}\ \bibnamefont
  {Pham}}, \bibinfo {author} {\bibfnamefont {M.}~\bibnamefont {Gabay}},\ and\
  \bibinfo {author} {\bibfnamefont {P.}~\bibnamefont {Lederer}},\ }\bibfield
  {title} {\bibinfo {title} {Fractional excitations in the luttinger liquid},\
  }\href@noop {} {\bibfield  {journal} {\bibinfo  {journal} {Physical Review
  B}\ }\textbf {\bibinfo {volume} {61}},\ \bibinfo {pages} {16397} (\bibinfo
  {year} {2000})}\BibitemShut {NoStop}%
\bibitem [{\citenamefont {Steinberg}\ \emph {et~al.}(2008)\citenamefont
  {Steinberg}, \citenamefont {Barak}, \citenamefont {Yacoby}, \citenamefont
  {Pfeiffer}, \citenamefont {West}, \citenamefont {Halperin},\ and\
  \citenamefont {Le~Hur}}]{Steinberg08}%
  \BibitemOpen
  \bibfield  {author} {\bibinfo {author} {\bibfnamefont {H.}~\bibnamefont
  {Steinberg}}, \bibinfo {author} {\bibfnamefont {G.}~\bibnamefont {Barak}},
  \bibinfo {author} {\bibfnamefont {A.}~\bibnamefont {Yacoby}}, \bibinfo
  {author} {\bibfnamefont {L.~N.}\ \bibnamefont {Pfeiffer}}, \bibinfo {author}
  {\bibfnamefont {K.~W.}\ \bibnamefont {West}}, \bibinfo {author}
  {\bibfnamefont {B.~I.}\ \bibnamefont {Halperin}},\ and\ \bibinfo {author}
  {\bibfnamefont {K.}~\bibnamefont {Le~Hur}},\ }\bibfield  {title} {\bibinfo
  {title} {Charge fractionalization in quantum wires},\ }\href@noop {}
  {\bibfield  {journal} {\bibinfo  {journal} {Nature Physics}\ }\textbf
  {\bibinfo {volume} {4}},\ \bibinfo {pages} {116} (\bibinfo {year}
  {2008})}\BibitemShut {NoStop}%
\bibitem [{\citenamefont {Hashisaka}\ \emph {et~al.}(2017)\citenamefont
  {Hashisaka}, \citenamefont {Hiyama}, \citenamefont {Akiho}, \citenamefont
  {Muraki},\ and\ \citenamefont {Fujisawa}}]{Hashisaka17}%
  \BibitemOpen
  \bibfield  {author} {\bibinfo {author} {\bibfnamefont {M.}~\bibnamefont
  {Hashisaka}}, \bibinfo {author} {\bibfnamefont {N.}~\bibnamefont {Hiyama}},
  \bibinfo {author} {\bibfnamefont {T.}~\bibnamefont {Akiho}}, \bibinfo
  {author} {\bibfnamefont {K.}~\bibnamefont {Muraki}},\ and\ \bibinfo {author}
  {\bibfnamefont {T.}~\bibnamefont {Fujisawa}},\ }\bibfield  {title} {\bibinfo
  {title} {Waveform measurement of charge-and spin-density wavepackets in a
  chiral tomonaga--luttinger liquid},\ }\href@noop {} {\bibfield  {journal}
  {\bibinfo  {journal} {Nature Physics}\ }\textbf {\bibinfo {volume} {13}},\
  \bibinfo {pages} {559} (\bibinfo {year} {2017})}\BibitemShut {NoStop}%
\bibitem [{\citenamefont {Giamarchi}(2003)}]{Giamarchi03}%
  \BibitemOpen
  \bibfield  {author} {\bibinfo {author} {\bibfnamefont {T.}~\bibnamefont
  {Giamarchi}},\ }\href@noop {} {\emph {\bibinfo {title} {Quantum physics in
  one dimension}}},\ Vol.\ \bibinfo {volume} {121}\ (\bibinfo  {publisher}
  {Clarendon press},\ \bibinfo {year} {2003})\BibitemShut {NoStop}%
\bibitem [{\citenamefont {Calzona}\ \emph {et~al.}(2015)\citenamefont
  {Calzona}, \citenamefont {Carrega}, \citenamefont {Dolcetto},\ and\
  \citenamefont {Sassetti}}]{Calzona15}%
  \BibitemOpen
  \bibfield  {author} {\bibinfo {author} {\bibfnamefont {A.}~\bibnamefont
  {Calzona}}, \bibinfo {author} {\bibfnamefont {M.}~\bibnamefont {Carrega}},
  \bibinfo {author} {\bibfnamefont {G.}~\bibnamefont {Dolcetto}},\ and\
  \bibinfo {author} {\bibfnamefont {M.}~\bibnamefont {Sassetti}},\ }\bibfield
  {title} {\bibinfo {title} {Transient dynamics of spin-polarized injection in
  helical luttinger liquids},\ }\href@noop {} {\bibfield  {journal} {\bibinfo
  {journal} {Physica E: Low-dimensional Systems and Nanostructures}\ }\textbf
  {\bibinfo {volume} {74}},\ \bibinfo {pages} {630} (\bibinfo {year}
  {2015})}\BibitemShut {NoStop}%
\bibitem [{\citenamefont {Perfetto}\ \emph {et~al.}(2014)\citenamefont
  {Perfetto}, \citenamefont {Stefanucci}, \citenamefont {Kamata},\ and\
  \citenamefont {Fujisawa}}]{Perfetto14_fracExpHall}%
  \BibitemOpen
  \bibfield  {author} {\bibinfo {author} {\bibfnamefont {E.}~\bibnamefont
  {Perfetto}}, \bibinfo {author} {\bibfnamefont {G.}~\bibnamefont
  {Stefanucci}}, \bibinfo {author} {\bibfnamefont {H.}~\bibnamefont {Kamata}},\
  and\ \bibinfo {author} {\bibfnamefont {T.}~\bibnamefont {Fujisawa}},\
  }\bibfield  {title} {\bibinfo {title} {Time-resolved charge fractionalization
  in inhomogeneous luttinger liquids},\ }\href@noop {} {\bibfield  {journal}
  {\bibinfo  {journal} {Physical Review B}\ }\textbf {\bibinfo {volume} {89}},\
  \bibinfo {pages} {201413(R)} (\bibinfo {year} {2014})}\BibitemShut {NoStop}%
\bibitem [{\citenamefont {Imambekov}\ \emph {et~al.}(2012)\citenamefont
  {Imambekov}, \citenamefont {Schmidt},\ and\ \citenamefont
  {Glazman}}]{Imambekov12}%
  \BibitemOpen
  \bibfield  {author} {\bibinfo {author} {\bibfnamefont {A.}~\bibnamefont
  {Imambekov}}, \bibinfo {author} {\bibfnamefont {T.~L.}\ \bibnamefont
  {Schmidt}},\ and\ \bibinfo {author} {\bibfnamefont {L.~I.}\ \bibnamefont
  {Glazman}},\ }\bibfield  {title} {\bibinfo {title} {One-dimensional quantum
  liquids: Beyond the luttinger liquid paradigm},\ }\href@noop {} {\bibfield
  {journal} {\bibinfo  {journal} {Reviews of Modern Physics}\ }\textbf
  {\bibinfo {volume} {84}},\ \bibinfo {pages} {1253} (\bibinfo {year}
  {2012})}\BibitemShut {NoStop}%
\bibitem [{\citenamefont {Rozhkov}(2005)}]{Rozhkov05}%
  \BibitemOpen
  \bibfield  {author} {\bibinfo {author} {\bibfnamefont {A.}~\bibnamefont
  {Rozhkov}},\ }\bibfield  {title} {\bibinfo {title} {Fermionic quasiparticle
  representation of tomonaga-luttinger hamiltonian},\ }\href@noop {} {\bibfield
   {journal} {\bibinfo  {journal} {The European Physical Journal B-Condensed
  Matter and Complex Systems}\ }\textbf {\bibinfo {volume} {47}},\ \bibinfo
  {pages} {193} (\bibinfo {year} {2005})}\BibitemShut {NoStop}%
\bibitem [{\citenamefont {Protopopov}\ \emph {et~al.}(2014)\citenamefont
  {Protopopov}, \citenamefont {Gutman}, \citenamefont {Oldenburg},\ and\
  \citenamefont {Mirlin}}]{Protopopov14}%
  \BibitemOpen
  \bibfield  {author} {\bibinfo {author} {\bibfnamefont {I.}~\bibnamefont
  {Protopopov}}, \bibinfo {author} {\bibfnamefont {D.}~\bibnamefont {Gutman}},
  \bibinfo {author} {\bibfnamefont {M.}~\bibnamefont {Oldenburg}},\ and\
  \bibinfo {author} {\bibfnamefont {A.}~\bibnamefont {Mirlin}},\ }\bibfield
  {title} {\bibinfo {title} {Dissipationless kinetics of one-dimensional
  interacting fermions},\ }\href@noop {} {\bibfield  {journal} {\bibinfo
  {journal} {Physical Review B}\ }\textbf {\bibinfo {volume} {89}},\ \bibinfo
  {pages} {161104(R)} (\bibinfo {year} {2014})}\BibitemShut {NoStop}%
\bibitem [{\citenamefont {Moreno}\ \emph {et~al.}(2013)\citenamefont {Moreno},
  \citenamefont {Muramatsu},\ and\ \citenamefont {Carmelo}}]{Moreno13}%
  \BibitemOpen
  \bibfield  {author} {\bibinfo {author} {\bibfnamefont {A.}~\bibnamefont
  {Moreno}}, \bibinfo {author} {\bibfnamefont {A.}~\bibnamefont {Muramatsu}},\
  and\ \bibinfo {author} {\bibfnamefont {J.~M.~P.}\ \bibnamefont {Carmelo}},\
  }\bibfield  {title} {\bibinfo {title} {Charge and spin fractionalization
  beyond the luttinger-liquid paradigm},\ }\href@noop {} {\bibfield  {journal}
  {\bibinfo  {journal} {Physical Review B}\ }\textbf {\bibinfo {volume} {87}},\
  \bibinfo {pages} {075101} (\bibinfo {year} {2013})}\BibitemShut {NoStop}%
\bibitem [{\citenamefont {Veness}\ and\ \citenamefont
  {Glazman}(2019)}]{Veness19}%
  \BibitemOpen
  \bibfield  {author} {\bibinfo {author} {\bibfnamefont {T.}~\bibnamefont
  {Veness}}\ and\ \bibinfo {author} {\bibfnamefont {L.~I.}\ \bibnamefont
  {Glazman}},\ }\bibfield  {title} {\bibinfo {title} {Fate of quantum shock
  waves at late times},\ }\href@noop {} {\bibfield  {journal} {\bibinfo
  {journal} {Physical Review B}\ }\textbf {\bibinfo {volume} {100}},\ \bibinfo
  {pages} {235125} (\bibinfo {year} {2019})}\BibitemShut {NoStop}%
\bibitem [{\citenamefont {Bettelheim}\ and\ \citenamefont
  {Glazman}(2012)}]{Bettelheim12}%
  \BibitemOpen
  \bibfield  {author} {\bibinfo {author} {\bibfnamefont {E.}~\bibnamefont
  {Bettelheim}}\ and\ \bibinfo {author} {\bibfnamefont {L.}~\bibnamefont
  {Glazman}},\ }\bibfield  {title} {\bibinfo {title} {Quantum ripples over a
  semiclassical shock},\ }\href@noop {} {\bibfield  {journal} {\bibinfo
  {journal} {Physical review letters}\ }\textbf {\bibinfo {volume} {109}},\
  \bibinfo {pages} {260602} (\bibinfo {year} {2012})}\BibitemShut {NoStop}%
\bibitem [{\citenamefont {Bettelheim}\ \emph {et~al.}(2006)\citenamefont
  {Bettelheim}, \citenamefont {Abanov},\ and\ \citenamefont
  {Wiegmann}}]{Bettelheim06}%
  \BibitemOpen
  \bibfield  {author} {\bibinfo {author} {\bibfnamefont {E.}~\bibnamefont
  {Bettelheim}}, \bibinfo {author} {\bibfnamefont {A.}~\bibnamefont {Abanov}},\
  and\ \bibinfo {author} {\bibfnamefont {P.}~\bibnamefont {Wiegmann}},\
  }\bibfield  {title} {\bibinfo {title} {Orthogonality catastrophe and shock
  waves in a nonequilibrium fermi gas},\ }\href@noop {} {\bibfield  {journal}
  {\bibinfo  {journal} {Physical review letters}\ }\textbf {\bibinfo {volume}
  {97}},\ \bibinfo {pages} {246402} (\bibinfo {year} {2006})}\BibitemShut
  {NoStop}%
\bibitem [{\citenamefont {Protopopov}\ \emph {et~al.}(2013)\citenamefont
  {Protopopov}, \citenamefont {Gutman}, \citenamefont {Schmitteckert},\ and\
  \citenamefont {Mirlin}}]{Protopopov13}%
  \BibitemOpen
  \bibfield  {author} {\bibinfo {author} {\bibfnamefont {I.}~\bibnamefont
  {Protopopov}}, \bibinfo {author} {\bibfnamefont {D.}~\bibnamefont {Gutman}},
  \bibinfo {author} {\bibfnamefont {P.}~\bibnamefont {Schmitteckert}},\ and\
  \bibinfo {author} {\bibfnamefont {A.}~\bibnamefont {Mirlin}},\ }\bibfield
  {title} {\bibinfo {title} {Dynamics of waves in one-dimensional electron
  systems: Density oscillations driven by population inversion},\ }\href@noop
  {} {\bibfield  {journal} {\bibinfo  {journal} {Physical Review B}\ }\textbf
  {\bibinfo {volume} {87}},\ \bibinfo {pages} {045112} (\bibinfo {year}
  {2013})}\BibitemShut {NoStop}%
\bibitem [{\citenamefont {Imambekov}\ and\ \citenamefont
  {Glazman}(2009)}]{Imambekov09}%
  \BibitemOpen
  \bibfield  {author} {\bibinfo {author} {\bibfnamefont {A.}~\bibnamefont
  {Imambekov}}\ and\ \bibinfo {author} {\bibfnamefont {L.~I.}\ \bibnamefont
  {Glazman}},\ }\bibfield  {title} {\bibinfo {title} {Universal theory of
  nonlinear luttinger liquids},\ }\href@noop {} {\bibfield  {journal} {\bibinfo
   {journal} {Science}\ }\textbf {\bibinfo {volume} {323}},\ \bibinfo {pages}
  {228} (\bibinfo {year} {2009})}\BibitemShut {NoStop}%
\bibitem [{\citenamefont {Von~Delft}\ and\ \citenamefont
  {Schoeller}(1998)}]{vonDelft98}%
  \BibitemOpen
  \bibfield  {author} {\bibinfo {author} {\bibfnamefont {J.}~\bibnamefont
  {Von~Delft}}\ and\ \bibinfo {author} {\bibfnamefont {H.}~\bibnamefont
  {Schoeller}},\ }\bibfield  {title} {\bibinfo {title} {Bosonization for
  beginners-refermionization for experts},\ }\href@noop {} {\bibfield
  {journal} {\bibinfo  {journal} {Annalen der Physik}\ }\textbf {\bibinfo
  {volume} {7}},\ \bibinfo {pages} {225} (\bibinfo {year} {1998})}\BibitemShut
  {NoStop}%
\bibitem [{\citenamefont {Pitaevskii}\ and\ \citenamefont
  {Lifshitz}(2012)}]{Pitaevskii2012}%
  \BibitemOpen
  \bibfield  {author} {\bibinfo {author} {\bibfnamefont {L.}~\bibnamefont
  {Pitaevskii}}\ and\ \bibinfo {author} {\bibfnamefont {E.}~\bibnamefont
  {Lifshitz}},\ }\href@noop {} {\emph {\bibinfo {title} {Physical Kinetics:
  Volume 10}}},\ Vol.~\bibinfo {volume} {10}\ (\bibinfo  {publisher}
  {Butterworth-Heinemann},\ \bibinfo {year} {2012})\BibitemShut {NoStop}%
\bibitem [{\citenamefont {Home}(2013)}]{home2013conceptual}%
  \BibitemOpen
  \bibfield  {author} {\bibinfo {author} {\bibfnamefont {D.}~\bibnamefont
  {Home}},\ }\bibfield  {title} {\bibinfo {title} {Conceptual foundations of
  quantum physics: an overview from modern perspectives},\ }\href@noop {} {\
  (\bibinfo {year} {2013})}\BibitemShut {NoStop}%
\bibitem [{\citenamefont {Davidson}\ \emph {et~al.}(2017)\citenamefont
  {Davidson}, \citenamefont {Sels},\ and\ \citenamefont
  {Polkovnikov}}]{davidson2017semiclassical}%
  \BibitemOpen
  \bibfield  {author} {\bibinfo {author} {\bibfnamefont {S.~M.}\ \bibnamefont
  {Davidson}}, \bibinfo {author} {\bibfnamefont {D.}~\bibnamefont {Sels}},\
  and\ \bibinfo {author} {\bibfnamefont {A.}~\bibnamefont {Polkovnikov}},\
  }\bibfield  {title} {\bibinfo {title} {Semiclassical approach to dynamics of
  interacting fermions},\ }\href@noop {} {\bibfield  {journal} {\bibinfo
  {journal} {Annals of Physics}\ }\textbf {\bibinfo {volume} {384}},\ \bibinfo
  {pages} {128} (\bibinfo {year} {2017})}\BibitemShut {NoStop}%
\bibitem [{\citenamefont {Shukla}\ and\ \citenamefont
  {Eliasson}(2011)}]{Shukla2011}%
  \BibitemOpen
  \bibfield  {author} {\bibinfo {author} {\bibfnamefont {P.}~\bibnamefont
  {Shukla}}\ and\ \bibinfo {author} {\bibfnamefont {B.}~\bibnamefont
  {Eliasson}},\ }\bibfield  {title} {\bibinfo {title} {Colloquium: Nonlinear
  collective interactions in quantum plasmas with degenerate electron fluids},\
  }\href@noop {} {\bibfield  {journal} {\bibinfo  {journal} {Reviews of Modern
  Physics}\ }\textbf {\bibinfo {volume} {83}},\ \bibinfo {pages} {885}
  (\bibinfo {year} {2011})}\BibitemShut {NoStop}%
\bibitem [{\citenamefont {Benedikter}\ \emph {et~al.}(2016)\citenamefont
  {Benedikter}, \citenamefont {Porta}, \citenamefont {Saffirio},\ and\
  \citenamefont {Schlein}}]{Benedikter2016}%
  \BibitemOpen
  \bibfield  {author} {\bibinfo {author} {\bibfnamefont {N.}~\bibnamefont
  {Benedikter}}, \bibinfo {author} {\bibfnamefont {M.}~\bibnamefont {Porta}},
  \bibinfo {author} {\bibfnamefont {C.}~\bibnamefont {Saffirio}},\ and\
  \bibinfo {author} {\bibfnamefont {B.}~\bibnamefont {Schlein}},\ }\bibfield
  {title} {\bibinfo {title} {From the hartree dynamics to the vlasov
  equation},\ }\href@noop {} {\bibfield  {journal} {\bibinfo  {journal}
  {Archive for Rational Mechanics and Analysis}\ }\textbf {\bibinfo {volume}
  {221}},\ \bibinfo {pages} {273} (\bibinfo {year} {2016})}\BibitemShut
  {NoStop}%
\bibitem [{\citenamefont {Benedikter}\ \emph {et~al.}(2015)\citenamefont
  {Benedikter}, \citenamefont {Porta},\ and\ \citenamefont
  {Schlein}}]{Benedikter2015}%
  \BibitemOpen
  \bibfield  {author} {\bibinfo {author} {\bibfnamefont {N.}~\bibnamefont
  {Benedikter}}, \bibinfo {author} {\bibfnamefont {M.}~\bibnamefont {Porta}},\
  and\ \bibinfo {author} {\bibfnamefont {B.}~\bibnamefont {Schlein}},\
  }\bibfield  {title} {\bibinfo {title} {Hartree-fock dynamics for weakly
  interacting fermions},\ }in\ \href@noop {} {\emph {\bibinfo {booktitle}
  {Mathematical Results in Quantum Mechanics: Proceedings of the QMath12
  Conference}}}\ (\bibinfo {organization} {World Scientific},\ \bibinfo {year}
  {2015})\ pp.\ \bibinfo {pages} {177--189}\BibitemShut {NoStop}%
\bibitem [{\citenamefont {Hepp}(1974)}]{Hepp1974}%
  \BibitemOpen
  \bibfield  {author} {\bibinfo {author} {\bibfnamefont {K.}~\bibnamefont
  {Hepp}},\ }\bibfield  {title} {\bibinfo {title} {The classical limit for
  quantum mechanical correlation functions},\ }\href@noop {} {\bibfield
  {journal} {\bibinfo  {journal} {Communications in Mathematical Physics}\
  }\textbf {\bibinfo {volume} {35}},\ \bibinfo {pages} {265} (\bibinfo {year}
  {1974})}\BibitemShut {NoStop}%
\bibitem [{\citenamefont {Brodin}\ \emph {et~al.}(2016)\citenamefont {Brodin},
  \citenamefont {Ekman},\ and\ \citenamefont {Zamanian}}]{brodin2016quantum}%
  \BibitemOpen
  \bibfield  {author} {\bibinfo {author} {\bibfnamefont {G.}~\bibnamefont
  {Brodin}}, \bibinfo {author} {\bibfnamefont {R.}~\bibnamefont {Ekman}},\ and\
  \bibinfo {author} {\bibfnamefont {J.}~\bibnamefont {Zamanian}},\ }\bibfield
  {title} {\bibinfo {title} {Quantum kinetic theories in degenerate plasmas},\
  }\href@noop {} {\bibfield  {journal} {\bibinfo  {journal} {Plasma Physics and
  Controlled Fusion}\ }\textbf {\bibinfo {volume} {59}},\ \bibinfo {pages}
  {014043} (\bibinfo {year} {2016})}\BibitemShut {NoStop}%
\bibitem [{\citenamefont {Gogolin}\ \emph {et~al.}(2004)\citenamefont
  {Gogolin}, \citenamefont {Nersesyan},\ and\ \citenamefont
  {Tsvelik}}]{Gogolin2004}%
  \BibitemOpen
  \bibfield  {author} {\bibinfo {author} {\bibfnamefont {A.~O.}\ \bibnamefont
  {Gogolin}}, \bibinfo {author} {\bibfnamefont {A.~A.}\ \bibnamefont
  {Nersesyan}},\ and\ \bibinfo {author} {\bibfnamefont {A.~M.}\ \bibnamefont
  {Tsvelik}},\ }\href@noop {} {\emph {\bibinfo {title} {Bosonization and
  strongly correlated systems}}}\ (\bibinfo  {publisher} {Cambridge university
  press},\ \bibinfo {year} {2004})\BibitemShut {NoStop}%
\end{thebibliography}%

\end{document}